\documentclass[preprint,prd,aps,showpacs,preprintnumbers,amsmath,amssymb]{revtex4-1}
\usepackage{graphics,epsfig,subfigure}
\usepackage{diagbox}
\usepackage[usenames]{color}
\usepackage{color}
\usepackage{graphicx}
\usepackage{amsfonts}
\usepackage{indentfirst}
\usepackage{booktabs}
\usepackage{hyperref}
\hypersetup{hypertex=ture,backref=true,colorlinks=ture,linkcolor=blue,anchorcolor=blue,citecolor=blue}
\usepackage{float}
\usepackage{multirow}

\usepackage{amsmath}

\begin{document}
\renewcommand{\baselinestretch}{1.3}
\newcommand\beq{\begin{equation}}
\newcommand\eeq{\end{equation}}
\newcommand\beqn{\begin{eqnarray}}
\newcommand\eeqn{\end{eqnarray}}
\newcommand\nn{\nonumber}
\newcommand\fc{\frac}
\newcommand\lt{\left}
\newcommand\rt{\right}
\newcommand\pt{\partial}

\title{\Large \bf Study of Boson Stars with Wormhole}
\author{Peng-Bo Ding, Tian-Xiang Ma, and Yong-Qiang Wang\footnote{yqwang@lzu.edu.cn, corresponding author
}
}

\affiliation{ $^{1}$Lanzhou Center for Theoretical Physics, Key Laboratory of Theoretical Physics of Gansu Province,
	School of Physical Science and Technology, Lanzhou University, Lanzhou 730000, China\\
	$^{2}$Institute of Theoretical Physics $\&$ Research Center of Gravitation, Lanzhou University, Lanzhou 730000, China}

\begin{abstract}
In this paper, we reconsider the mixed system of BSs with wormholes at their center which performed by complex scalar field and phantom field and study a whole new condition about the potential. Both the symmetric and asymmetric solutions in the two asymptotically flat regions are obtained by using numerical method and we mainly explore the change of the results by varying the parameters of throats and potential. In ground state, we find there are multiple solutions at certain setting of parameters and with the increase of $\eta_0$ or decrease of $c$, the results gradually become single-valued functions and these two variables have similar influence to the curve shape of mass $M$ and charge $Q$, furthermore, the asymmetric solutions can turn into the solutions of symmetry at some frequency $\omega$ in certain $\eta_0$ and $c$. However, when it comes to excited state, the properties of solutions of symmetry is similar to the ground state while asymmetrical results exhibit altered characteristics. We also present the geometries of wormhole to investigate the property of this model.
\end{abstract}

\maketitle

\section{Introduction}\label{Sec1}

Aiming to seek stable, particle-like entities which can span the extent from fundamental to cosmic, in 1955 and 1957, J. A. Wheeler coupled the classical fields of electromagnetism with general relativity. However, he got an unstable solution and named such source-free objects \textit{geons}~\cite{Wheeler:1955zz,Power:1957zz}. After a decade, the electromagnetism theory was superseded by D. J. Kaup's complex scalar field approach and he found the solution of this configuration was stable~\cite{Kaup:1968zz}, so-called \textit{Klein-Gordon geons}. Then, R. Ruffini studied the quantized real scalar field coupled to Einstein's gravity and found the similar equations\cite{Ruffini:1969qy}. All of these diverse forms have become familiar as present \textit{Boson stars} (BSs)\cite{Mielke:1996a}.

Wormholes, which are fascinating astrophysical objects, have been the target of observational searches\cite{Abe:2010ap,Toki:2011zu,Takahashi:2013jqa}, and on theoretical side, in dilatonic Einstein-Gauss-Bonnet theory, traversable wormholes in four-dimensional spacetimes have been considered in Refs.\cite{Kanti:2011jz,Kanti:2011yv}. What's more, in the mixed system, lorentzian wormholes with chiral matter fields and a phantom field have been investigated\cite{Charalampidis:2013ixa}. A short time ago, rotating neutron stars which have wormholes in their centers have been addressed\cite{Dzhunushaliev:2022elv}.

Eills found Einstein's gravity indicate the existence of nontrivial topology wormholes and mixed systems requires a violation of the energy conditions, and can be expressed as a phantom field\cite{Ellis:1973yv,Ellis:1979bh,Bronnikov:1973fh,Morris:1988cz,Morris:1988tu,Lobo:2005us}. And phantom's existence might become a fact because \textit{dark energy} is the primary component of the universe in nowadays. Moreover, there are many suggestions that should involve scalar fields to support the theory of dark matter, and BSs can be the prime candidates\cite{Sahni:1999qe,Matos:2000ng,Hu:2000ke,Suarez:2013iw,Hui:2016ltb,Padilla:2019fju}.

BSs with trivial topology has been the subject of extensive research. Such as BSs with self-interaction\cite{Colpi:1986ye,Mielke:1980sa,Kling:2017hjm,Herdeiro:2020jzx}, the Newtonian boson stars, which are the solution of Einstein's Klein-Gordon equation\cite{harrison2002numerical}. And we are interested in nontrivial topology construction, by considering the mixed systems that consist BSs with wormholes at their center we can get such configuration. In addition, as one of the solution of field equation, Q-balls can be an excellent entry point. Most recently, there are some related research\cite{Dzhunushaliev:2014bya,Hoffmann:2017jfs}.

Also, we are interested in the configuration of other BSs coupled to wormholes. There are a study about ground state has been conducted by Hoffmann\cite{Hoffmann:2018oml}. In this paper, we further investigate both symmetric and asymmetrical solution of BSs in this mixed system and describe it by numerical method. In particular, we meticulously study the conversion of the system's properties by changing the parameters of the potential and wormhole. Besides, we explore the wormhole geometries and map the situation of single throat and double throat. We study the excited state and analyze the characteristics as well.

This paper is structured in the following manner. In Sec.~\ref{sec2}, we describe the theoretical model about wormhole configurations that are embedded in bosonic matter in four dimensions. In Sec.~\ref{sec3}, the boundary conditions of these configurations are provided. We exhibit the numerical results of ground state in Sec.~\ref{sec4}, mainly concerning the mass and Noether charge. Ulteriorly, in Sec.~\ref{sec5}, The condition of excited state is explored. We make a summary in Sec.~\ref{sec6}.

\section{The model}\label{sec2}

\subsection{Action}

Considering a phantom field $\Psi$ and a scalar field $\Phi$ in general relativity, the corresponding Einstein-Hilbert action is
\begin{equation}\label{equ1}
S=\int\left[\frac{1}{2 \kappa} R+\mathcal{L}_{\mathrm{ph}}+\mathcal{L}_{\mathrm{bs}}\right] \sqrt{-g} d^4 x\,,
\end{equation}
\noindent there are two related Lagrangians, the Lagrangians $\mathcal{L}_{\mathrm{bs}}$ about complex scalar field $\Phi$
\begin{equation}
\mathcal{L}_{\mathrm{bs}}=-\frac{1}{2} g^{\mu \nu}\left(\partial_\mu \Phi^* \partial_\nu \Phi+\partial_\nu \Phi^* \partial_\mu \Phi\right)-U(|\Phi|)\,,
\end{equation}
the term $\mathcal{L}_{\mathrm{ph}}$ about phantom field $\Psi$
\begin{equation}
\mathcal{L}_{\mathrm{ph}}=\frac{1}{2} \partial_\mu \Psi \partial^\mu \Psi\,,
\end{equation}
 \noindent the symbol * denotes the operation of complex conjugation, the potential U can be written in the form
\begin{equation}
U(|\Phi|)=|\Phi|^2\left(a|\Phi|^4-c|\Phi|^2+b\right)\,,
\end{equation}
 \noindent b can be regard as mass term, and when the parameter a and c is positive, we can get solution of Q-balls\cite{Coleman:1985ki}, therefore, in this paper we choose negative number or zero as the value of c and $a=0$ in order to obtain other soliton solutions of non-trivial topology.

 Varying the action (\ref{equ1}) with respect to the metric tensor lead to the Einstein equations of the mixed system
\begin{equation}\label{equ5}
R_{\mu \nu}-\frac{1}{2} g_{\mu \nu} R=\kappa T_{\mu \nu}\,,
\end{equation}
and stress-energy tensor
\begin{equation}
T_{\mu \nu}=g_{\mu \nu} (\mathcal{L}_{ph}+\mathcal{L}_{bs})-2 \frac{\partial (\mathcal{L}_{ph}+\mathcal{L}_{bs})}{\partial g^{\mu \nu}}\,,
\end{equation}

\subsection{Ansatz}

Due to we focus on the spherically symmetric configuration, the ansatz can be chosen as following form\cite{Hoffmann:2017jfs}:
\begin{equation}
d s^2=-e^{- A} d t^2+p e^{-A}\left[d \eta^2+\left(\eta^2+\eta_0^2\right) d\Omega^2\right]\,,
\end{equation}
where $A$ and $p$ are functions of $\eta$, $\eta_0$ is the throat parameter. The $\eta$ is the radial coordinate, its value range is all real numbers, positive and negative value represent two independent universes.

\subsection{Einstein field equation}

Because of the geometry is spherically symmetric, we can parametrize the complex scalar field $\Phi$ and phantom field $\Psi$, and reads:
\begin{equation}
\Phi=\phi(\eta)e^{i \omega t}\,,
\end{equation}
\begin{equation}
\Psi=\psi(\eta)\,.
\end{equation}
Here $\omega$ denote the frequency, both $\phi(\eta)$ and $\psi(\eta)$ are real functions.

Substituting the aforementioned ansatz into Einstein equation (\ref{equ5}),
\begin{equation}\label{equ10}
\begin{aligned}
& -\frac{1}{p} e^A\left(A^{\prime \prime}-p^{\prime \prime} / p\right)+\frac{1}{4p} e^A\left(A^{\prime 2}-2 A^{\prime} p^{\prime} / p\right) \\
& -\frac{3}{4 p^3} e^A p^{\prime 2}-\frac{2}{p h} e^A \eta\left(A^{\prime}-p^{\prime} / p\right) \\
& +\frac{1}{p h^2} e^A \eta_0^2=-\kappa\left(U(\phi)+\frac{e^A}{2 p}\left(2 \phi^{\prime 2}-\psi^{\prime 2}\right)+\omega_s^2 e^{- A} \phi^2\right)\,,
\end{aligned}
\end{equation}

\begin{equation}\label{equ11}
\begin{aligned}
& -\frac{1}{4 p} e^A A^{\prime 2}+\frac{1}{4 p^3} e^A p^{\prime 2} \\
& +\frac{1}{ p^2 h} e^A \eta p^{\prime}-\frac{1}{ p h^2} e^A \eta_0^2=-\kappa\left(U(\phi)-\frac{e^A}{2 p}\left(2 \phi^{\prime 2}-\psi^{\prime 2}\right)-\omega_s^2 e^{- A} \phi^2\right)\,,
\end{aligned}
\end{equation}

\begin{equation}\label{equ12}
\begin{aligned}
& \frac{1}{2 p^2} e^A p^{\prime \prime}+ \frac{1}{4 p} e^A A^{\prime 2} \\
&-\frac{1}{2 p^3} e^A p^{\prime 2}+\frac{1}{2 p^2 h} e^A \eta p^{\prime} \\
&+\frac{1}{ p h^2} e^A \eta_0^2=-\kappa\left(U(\phi)+\frac{e^A}{2 p}\left(2 \phi^{\prime 2}-\psi^{\prime 2}\right)-\omega_s^2 e^{- A} \phi^2\right)\,,
\end{aligned}
\end{equation}
arising from the $tt$, $\eta\eta$, $\theta\theta$ components, respectively, where $h=\eta^2+\eta_0^2$

Varying the action (\ref{equ1}) with respect to the phantom field and to the complex scalar field lead to the field equations
\begin{equation}\label{equ13}
\left(p h \frac{1}{\sqrt{p}} \psi^{\prime}\right)^{\prime}  =0\,,
\end{equation}

\begin{equation}\label{equ14}
\phi^{\prime \prime}+\left( \frac{p^{\prime}}{2p}+ \frac{2\eta}{h}\right) \phi^{\prime}  =\frac{1}{2} p e^{-A} \frac{d U}{d \phi}-\omega_s^2 p e^{-2 A} \phi\,.
\end{equation}
Integrating the equation (\ref{equ13}):
\begin{equation}\label{equ15}
\psi^{\prime}=\sqrt{p}(p h)^{-1} \mathcal{D}\,,
\end{equation}
the constant D is the scalar charge of the phantom field, and substituting the Eq. (\ref{equ15}) into Einstein equations (\ref{equ10}-\ref{equ12}), the term of phantom field $\psi'^2$ can be excluded.

With the goal of solving the functions of metric, we need to eliminate the $\psi'^2$ and $\phi'^2$, by adding Eq. (\ref{equ11}) to Eq. (\ref{equ10}) and to Eq. (\ref{equ12}), $A(\eta)$ and $p(\eta)$ can be read:
\begin{equation}
A^{\prime \prime}+\frac{2}{h}\eta A^{\prime}-\frac{1}{2 p}  A^{\prime} p^{\prime} =-2 \kappa p e^{-A}\left(U(\phi)-2\omega_s^2 \phi^2 e^{- A}\right)\,,
\end{equation}
\begin{equation}
p^{\prime \prime}+\frac{3}{h} \eta p^{\prime}-\frac{1}{2 p} p^{\prime 2}=-4 \kappa p^2 e^{-A}\left(U(\phi)-\omega_s^2 \phi^2 e^{-A}\right)\,,
\end{equation}
then, along with the Eq. (\ref{equ14}), we get a system of ODEs that can be solved numerically.

\subsection{Charge and mass}

The action is invariant under the transformation $\phi \rightarrow \phi e^{i\alpha}$ with $\alpha$ is arbitrary  constant, according to Noether's theory, it lead to the conserved current
\begin{equation}
j^\mu=-i\left(\Phi^* \partial^\mu \Phi-\Phi \partial^\mu \Phi^*\right), \quad j_{; \mu}^\mu=0 .
\end{equation}
Integrating the component of time, Noether charge can be read:
\begin{equation}\label{19}
Q=-\int j^t|g|^{1 / 2} d \eta d \Omega=\int \frac{8\pi \omega \phi^2}{e^{2A}}h p^{3/2}d\eta \,.
\end{equation}
The mass $M$ of stationary, asymptotically flat solutions can be derived using the Komar expression\cite{Wald:1984rg}
\begin{equation}\label{20}
M=\frac{1}{4 \pi} \int_{\Sigma} R_{\mu \nu} n^\mu \xi^\nu d V \,,
\end{equation}
$\Sigma$ is used to represent a spacelike hypersurface that approaches asymptotic flatness, $dV$ represents the volume element on $\Sigma$.

By integrating Eq. (\ref{19}) and Eq. (\ref{20}) from $\eta=0$ to $\pm \infty$, we can obtain the charge $Q_\pm$ and the mass $M_\pm$ and we remain true in the numerical result.

\section{Boundary conditions}\label{sec3}
We need six boundary condition for solving the above three connected ODEs, aiming to prevent energy from becoming divergent, the scalar field function $\psi(\eta)$ should abide
\begin{equation}
\phi (\pm \infty) = 0\,,
\end{equation}
For the situation of $\eta \rightarrow \pm \infty$, the spacetime is supposed to the asymptotic flatness. Therefore, the metric function $A(\eta)$ obey
\begin{equation}
A (\pm \infty) = 0\,,
\end{equation}
same rules for metric function $p(\eta)$
\begin{equation}
p (\pm \infty) = 1\,,
\end{equation}
although these boundary conditions are imposed from symmetric solution, asymmetric condition can also arise naturally.

\section{Ground states}\label{sec4}
Here we study the symmetric and asymmetric solutions of ground state and discuss the connection of the two models. Especially, we focus on the transformation of charge $Q_\pm$ and mass $M_\pm$ when we vary the charge $\eta_0, c, \omega$.

To assist numerical calculations, we introduce a new radial coordinate x
\begin{equation}
\eta=tan(x\frac{\pi}{2})\,,
\end{equation}
$\eta \in (-\infty,+\infty)$ corresponding to $x \in (-1,1)$. We use finite element algorithm of Ritz-Galerkin to numerically solve the ODEs. In symmetrical condition, the region $-1\geq x \leq 1$ is discretized into a grid of 4000 points, and in some condition of asymmetry, the number of points is increased to 10000. Additionally, the relative error to be less than $10^{-4}$ to ensure the correctness of result. The parameter $\kappa$ is fixed to $0.2$.

When the parameters $\eta_0$ and $c$ are selected for some values, the solutions exhibit $multi-branch$, which means for some certain frequency $\omega$, metric function $A,p$ and complex scalar field function $\phi$ will be more than one result. Next, the $one-branch$ solutions and $multi-branch$ solutions will be employed. We call the branch of containing the result "$\omega=1, M=Q=0$" is the first branch, and if there are two branch contain that result, the one with the larger remaining $M, Q$ is the first branch, the branch adjacent to the first branch is called the second branch and so on.

\subsection{Symmetric results}

We divide our symmetry results into two categories: $c=0$ correspond to the potential only possesses mass term and $c\neq 0$ correspond to the potential obtain besides mass term, also a quartic term. For $c=0$, we change the value of $\eta_0$ to investigate the difference of solutions, and for $c\neq 0$, the parameter of wormhole is fixed at $\eta_0=0.5$, we vary the value of $c$ and discuss the transformation of results.

\begin{figure}[!htbp]
\begin{center}
\includegraphics[height=.17\textheight]{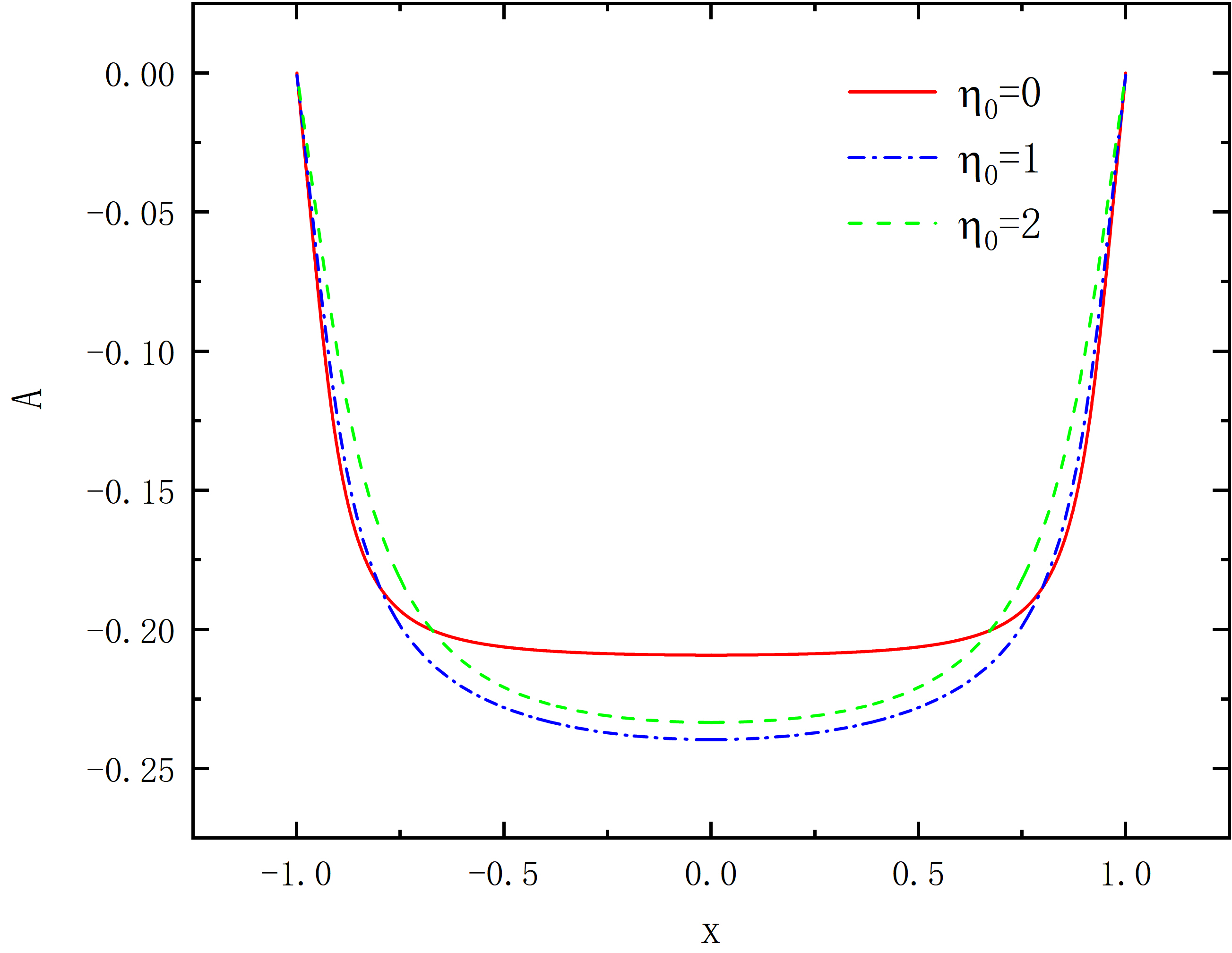}
\includegraphics[height=.17\textheight]{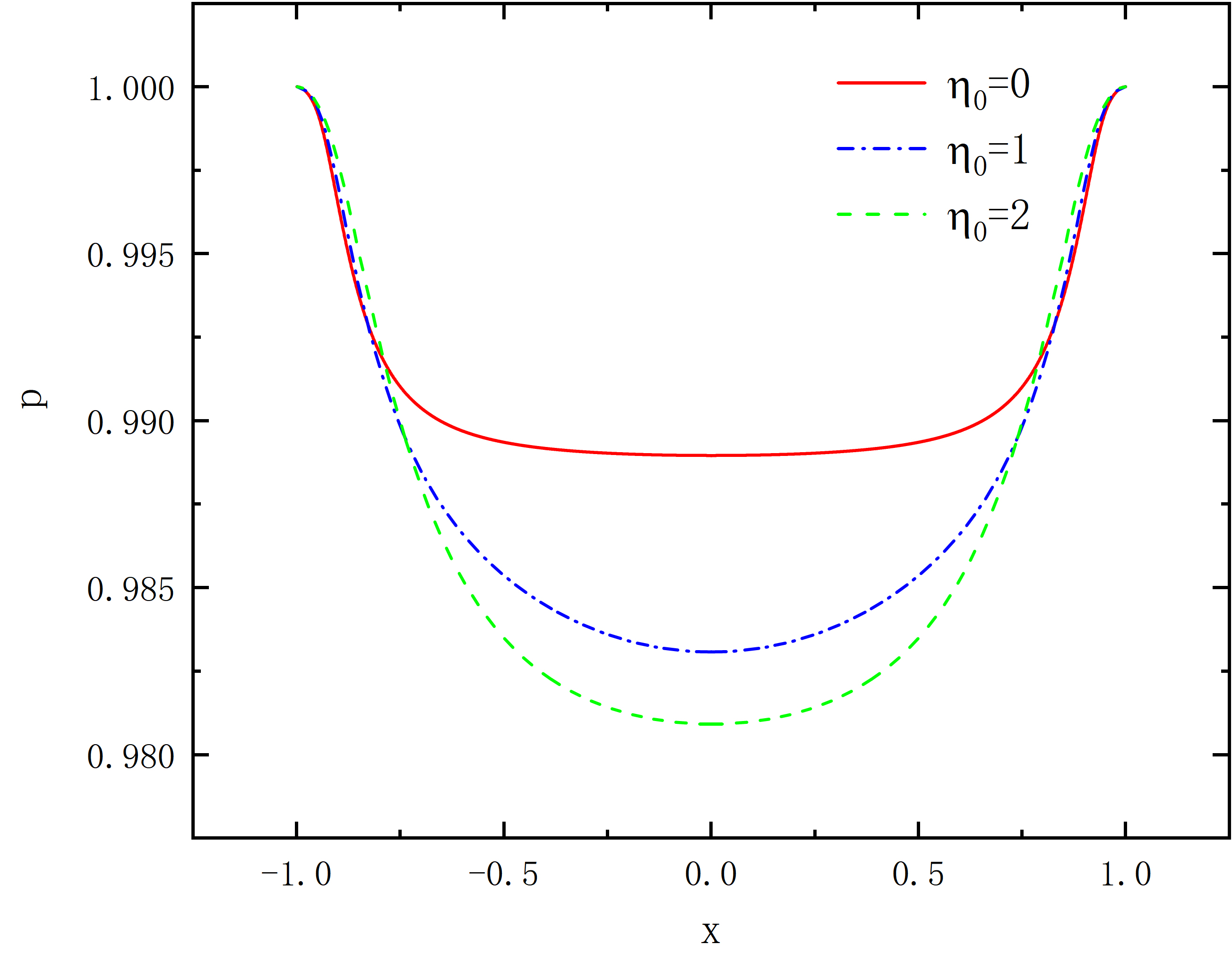}
\includegraphics[height=.17\textheight]{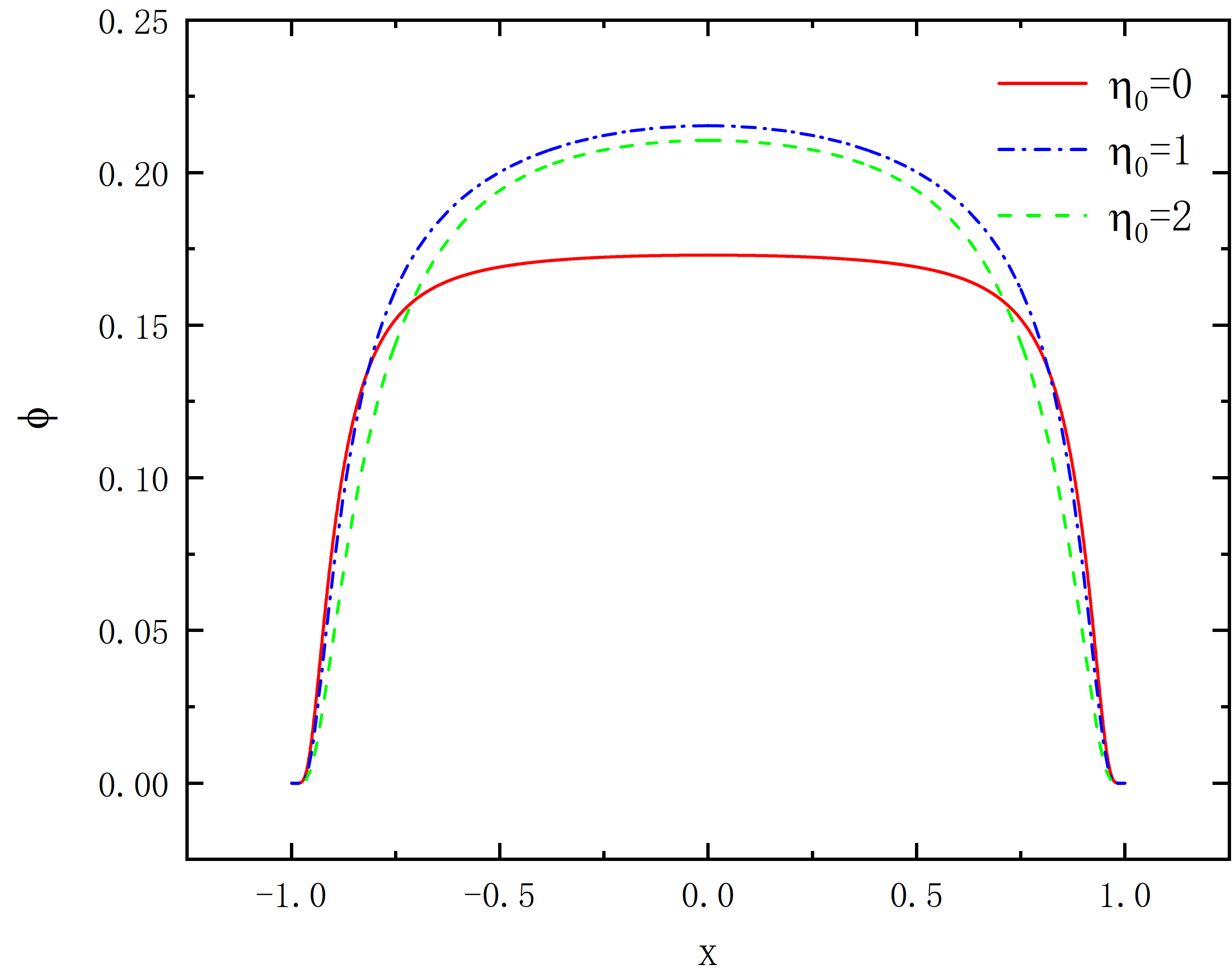}
\end{center}
\caption{Metric function $A$ (left panel) and $p$ (middle panel) with scalar field $\phi$ (right panel) as functions of x for $\eta_0=0,1,2$ in the case of symmetry, and $c=0,\omega=0.95$.}
\label{Sphi_A_p}
\end{figure}

The numerical result of the scalar field function and metric function at $\omega =0.95$ are shown in Fig. \ref{Sphi_A_p}, all solutions are in the first branch or only have one branch. We can find when $\eta_0=0$, these three functions all perform flat in the region of $x$ near zero, and as the $\eta_0$ increases, the region become smaller and the absolute value of gradient near $x=\pm 1$ are bigger.

\begin{figure}[!htbp]
\begin{center}
\includegraphics[height=.26\textheight]{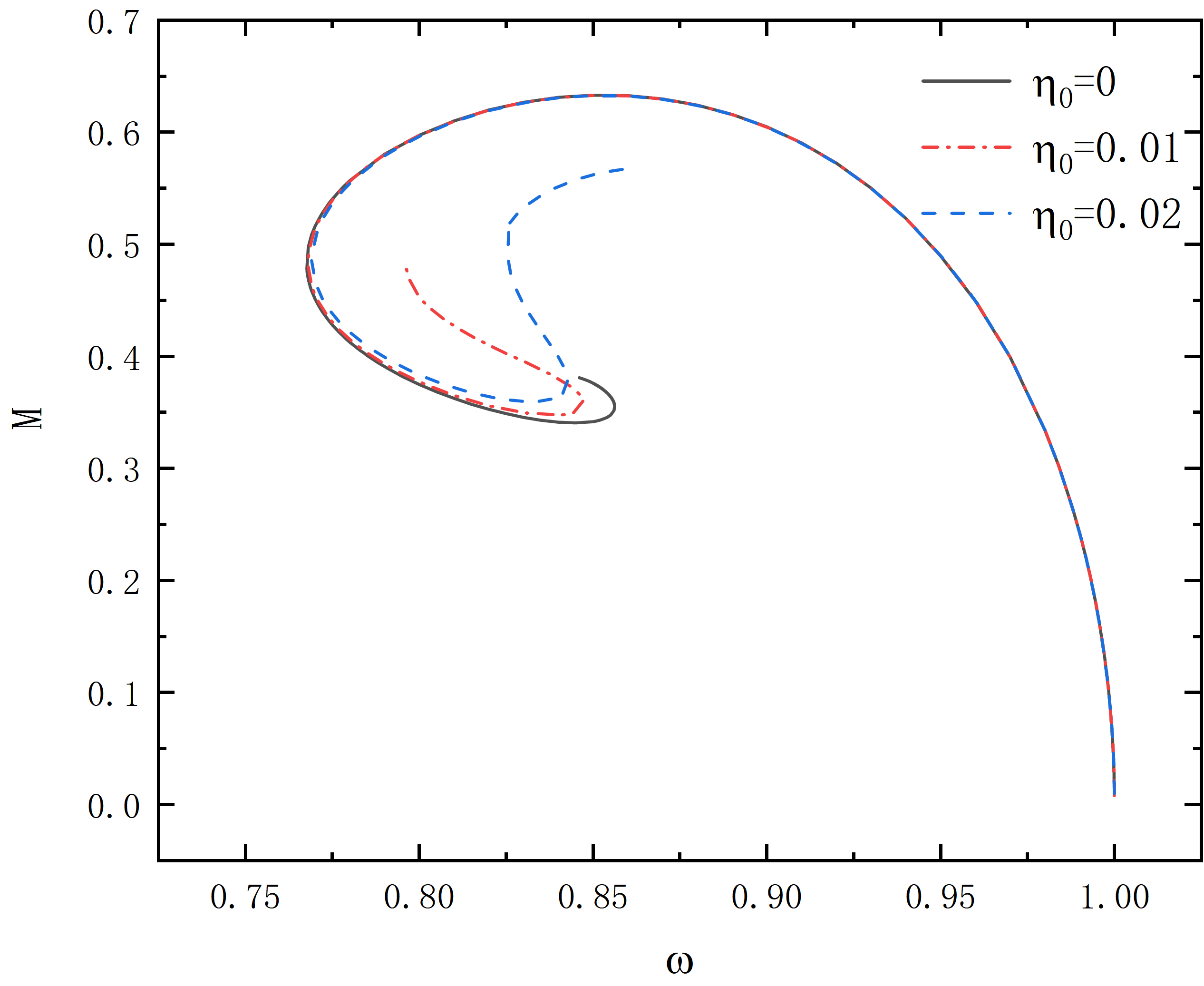}
\includegraphics[height=.26\textheight]{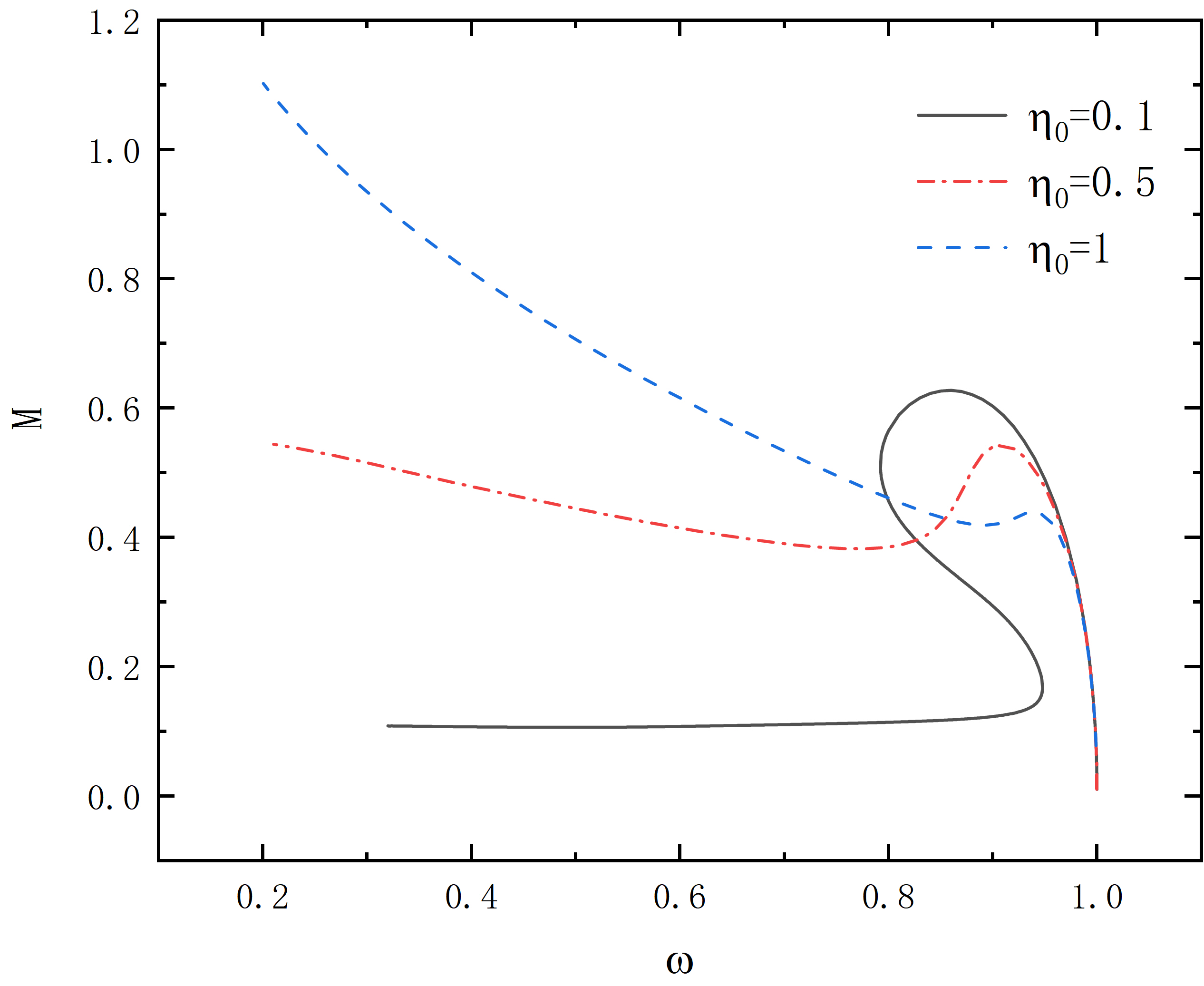}
\includegraphics[height=.26\textheight]{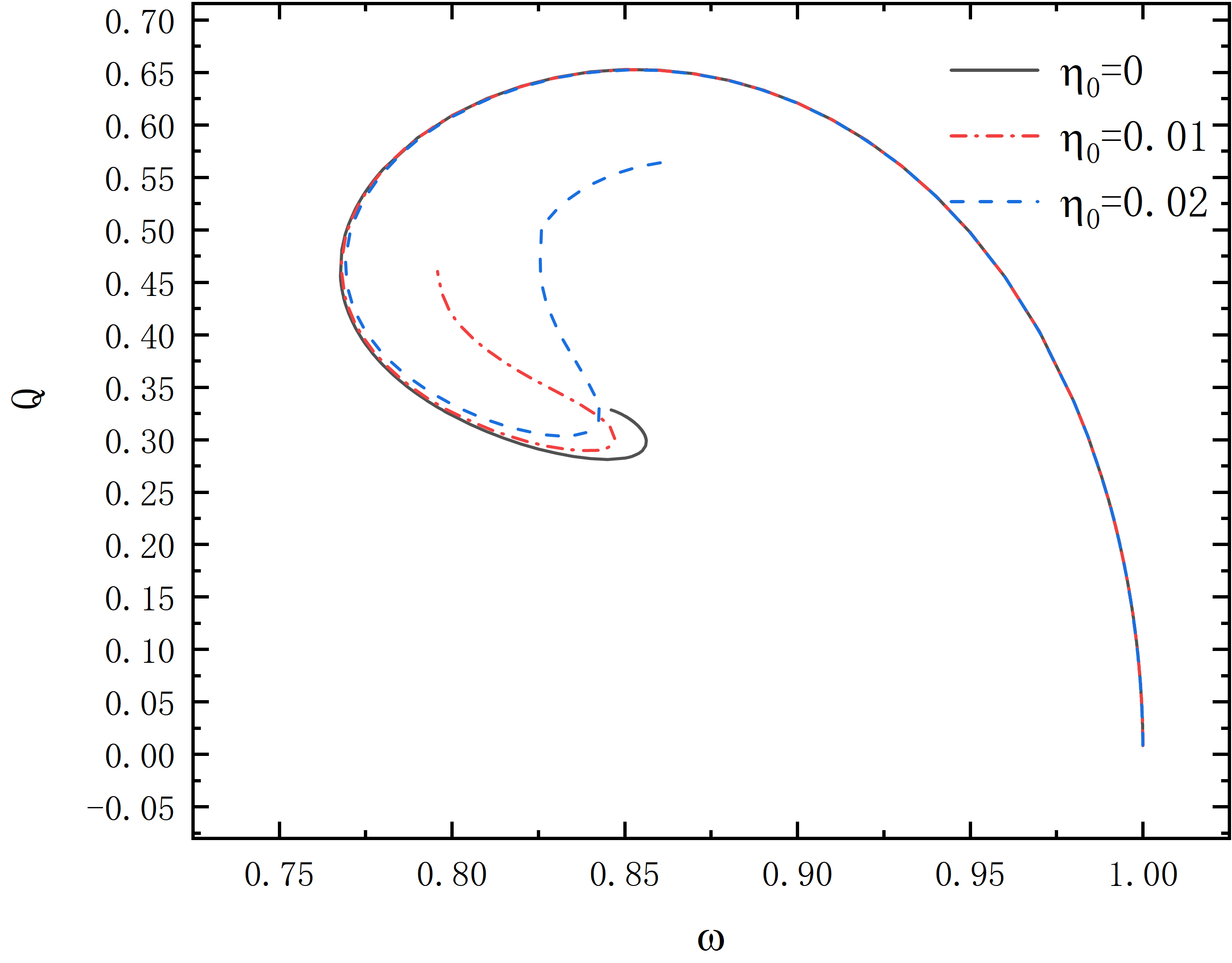}
\includegraphics[height=.26\textheight]{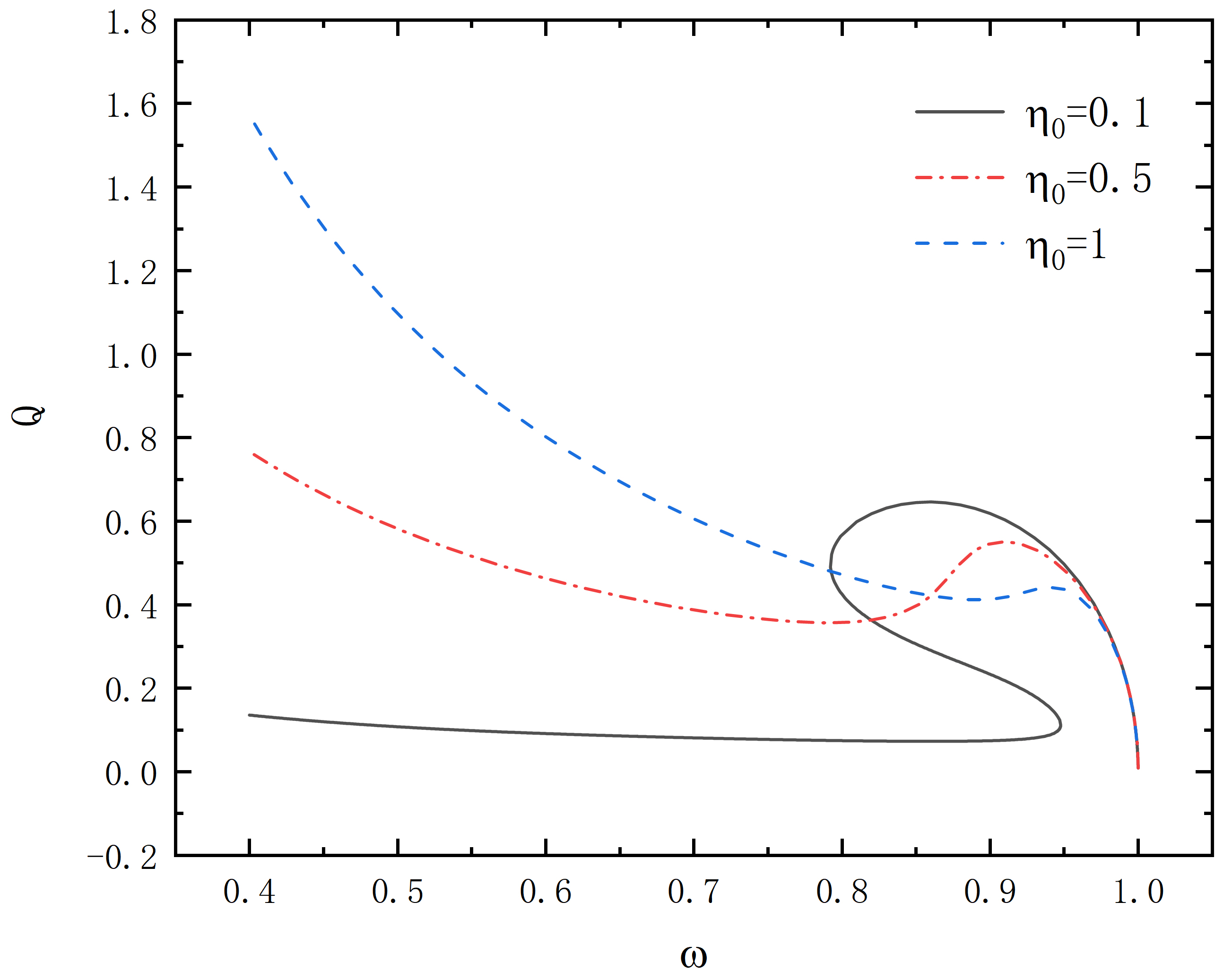}
\end{center}
\caption{The mass $M$ and the charge $Q$ as functions of frequency $\omega$ for $\eta_0=0,0.01,0.02,0.1,0.5,1$ in the case of symmetry, all solutions have $c=0$.}
\label{0SMQ}
\end{figure}
In Fig. \ref{0SMQ}, we exhibit the value of mass and charge under different $\omega$ at the situation of $c=0$. Because of the solutions are symmetric, $M_+=M_-, Q_+=Q_-$, we merely calculate the $M_+$ and $Q_+$ and represent with $M, Q$ in the figure. The data of $\eta_0=0$ corresponds to the condition without phantom field. Moreover, the curve shape of Mass $M$ is very similar to the charge $Q$ when $\omega$ is close to one.

Particularly, if $\eta_0$ is small, the results will be limited in a tight range of $\omega$, the solutions will become $multi-branch$ solutions, and curve shapes are just like to the solutions of isolated scalar field. As the value of $\eta_0$ goes up, the branch of high degree and the branch of low degree gradually combine together and then the branch of low degree merge with the branch of lower degree, in the end, the outcome only have one branch, such as the result of $\eta_0=1$, also the domain of functions is expanded, and he data will be divergent at $\omega =0$.

\begin{figure}[!htbp]
\begin{center}
\includegraphics[height=.26\textheight]{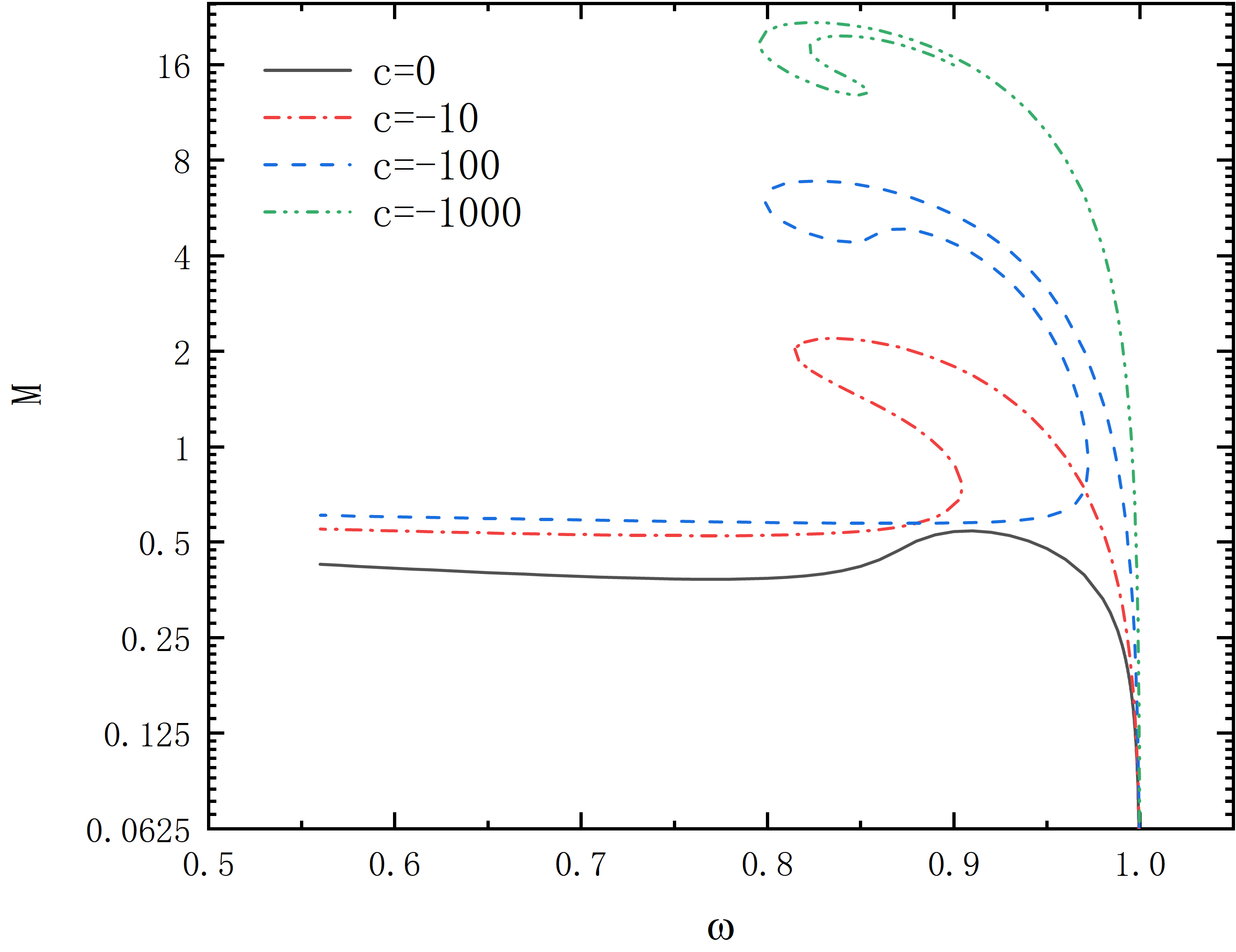}
\includegraphics[height=.26\textheight]{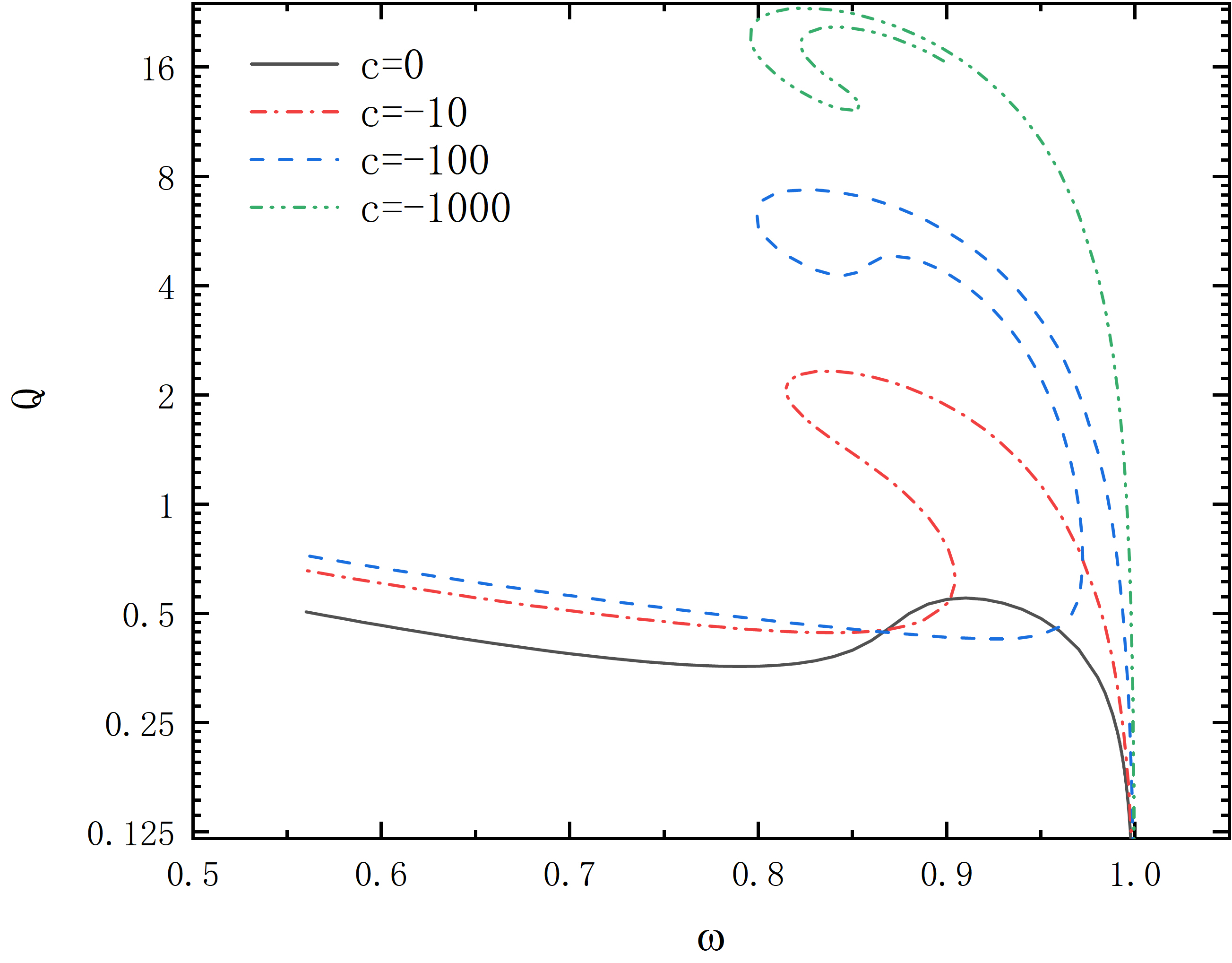}
\end{center}
\caption{The mass $M$ and the charge $Q$ as functions of frequency $\omega$ for $c=0,-10,-100,-1000$ in the case of symmetry, all solutions have $\eta_0=0.5$.}
\label{CSMQ}
\end{figure}

When it comes to the solutions of the potential obtain a quartic term, as shown in Fig. \ref{CSMQ}, the black line corresponds to the situation of $c=0$, which can also seen in Fig. \ref{0SMQ}. The graph provides a direct visualization of the rise in the $M$ and $Q$ at equivalent $\omega$ accompanied by the growth of absolute value of $c$. Also, it presents the trend of emerging branch in the setting of high value of the quartic term in the potential and the curve shape are similar to the solutions of low value of $\eta_0$. For example, the red line corresponds to the $\eta_0=1$ in Fig. \ref{0SMQ}, that indicates the influence of the quartic term is like the affect of the $\eta_0$ in the study of branch.

\subsection{Asymmetric results}

Similar to the case of symmetry, we divide the asymmetry results into two categories: $c=0$ and $c\neq 0$. Furthermore, the transformation between the symmetric solutions and the asymmetric solutions is explored.

\begin{figure}[!htbp]
\begin{center}
\includegraphics[height=.17\textheight]{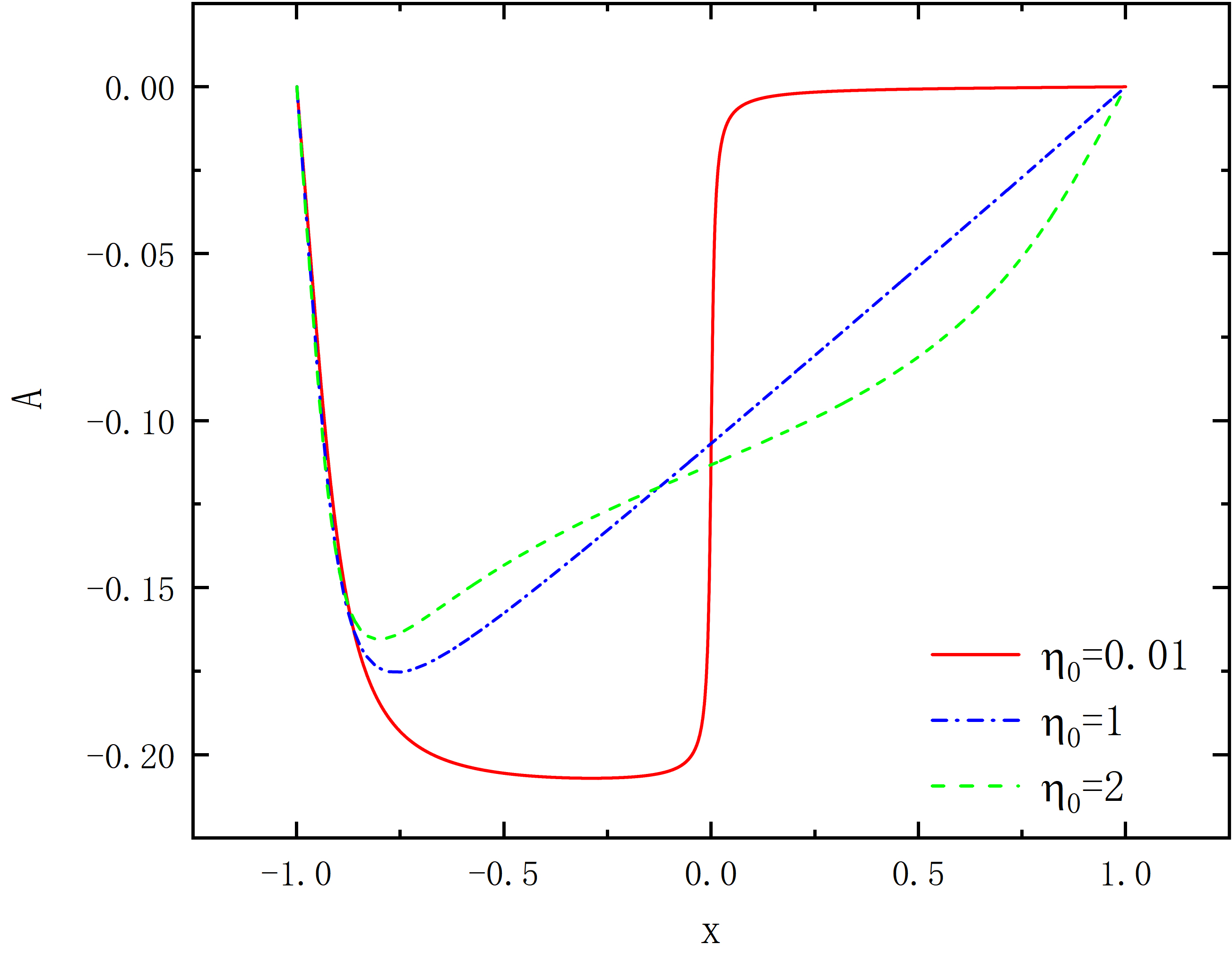}
\includegraphics[height=.17\textheight]{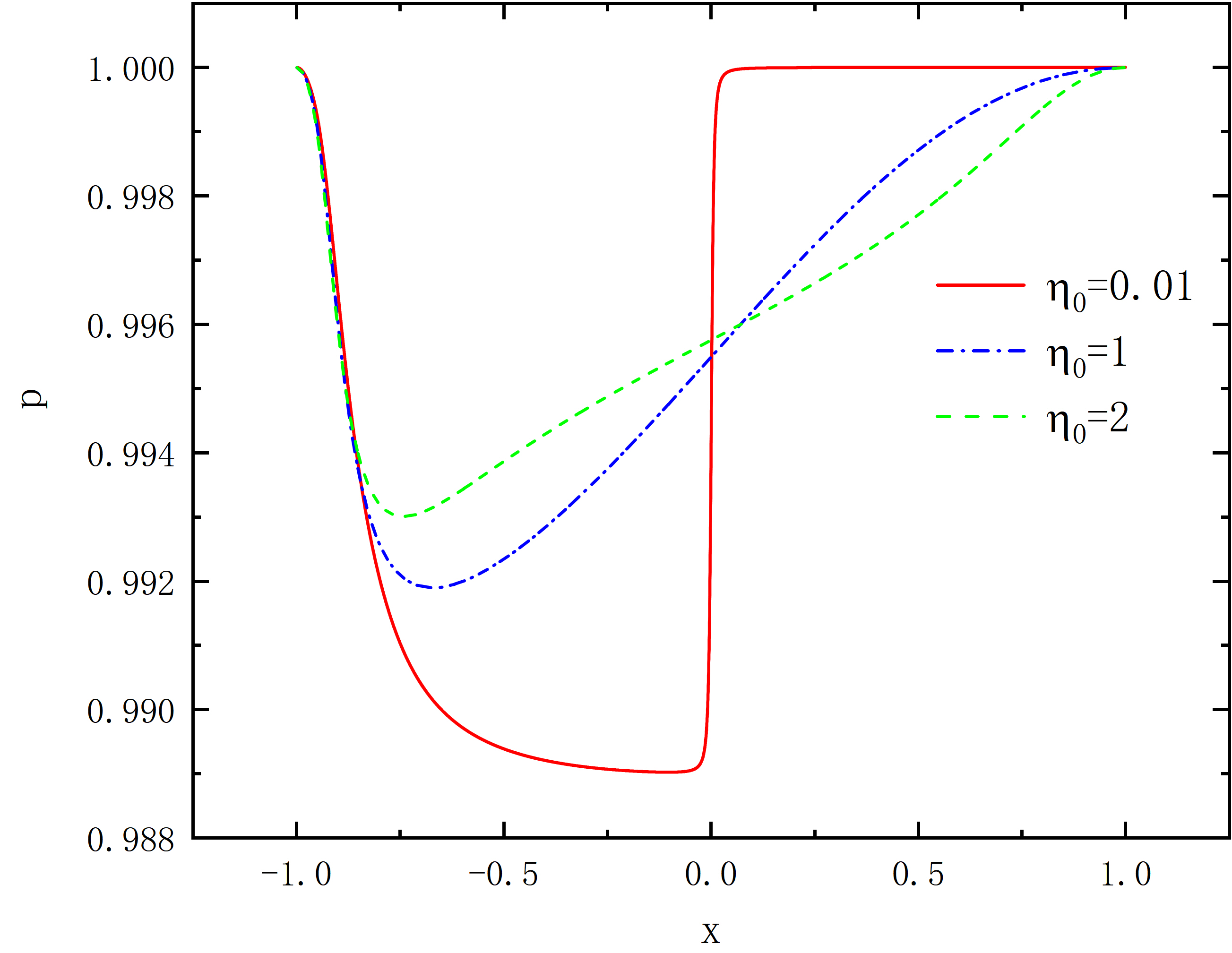}
\includegraphics[height=.17\textheight]{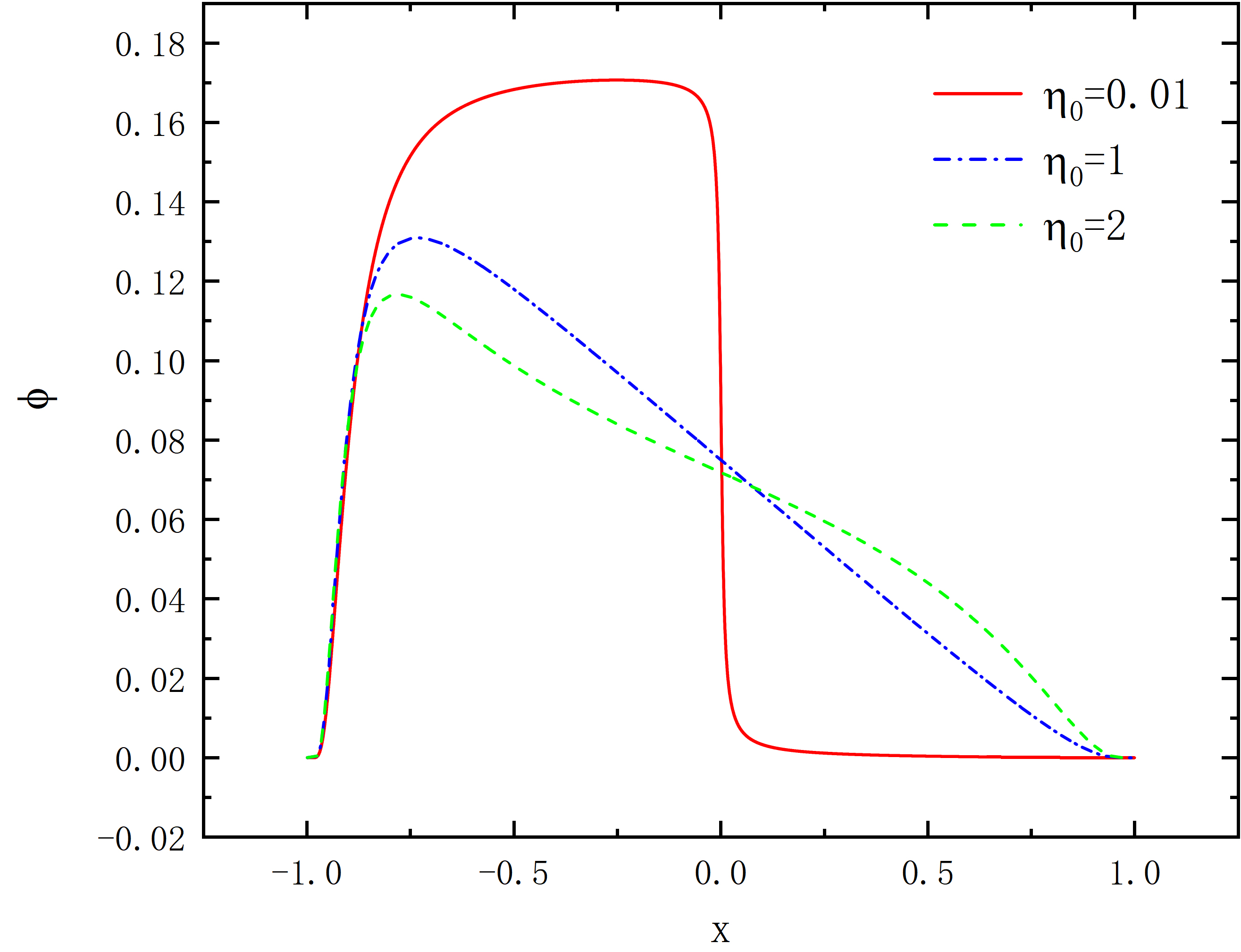}
\end{center}
\caption{Metric function $A$ (left panel) and $p$ (middle panel) with scalar field $\phi$ (right panel) as functions of x for $\eta_0=0,1,2$ in the case of asymmetry, and $c=0,\omega=0.95$.}
\label{AS_phi_A_p}
\end{figure}

\begin{figure}[!htbp]
\begin{center}
\includegraphics[height=.26\textheight]{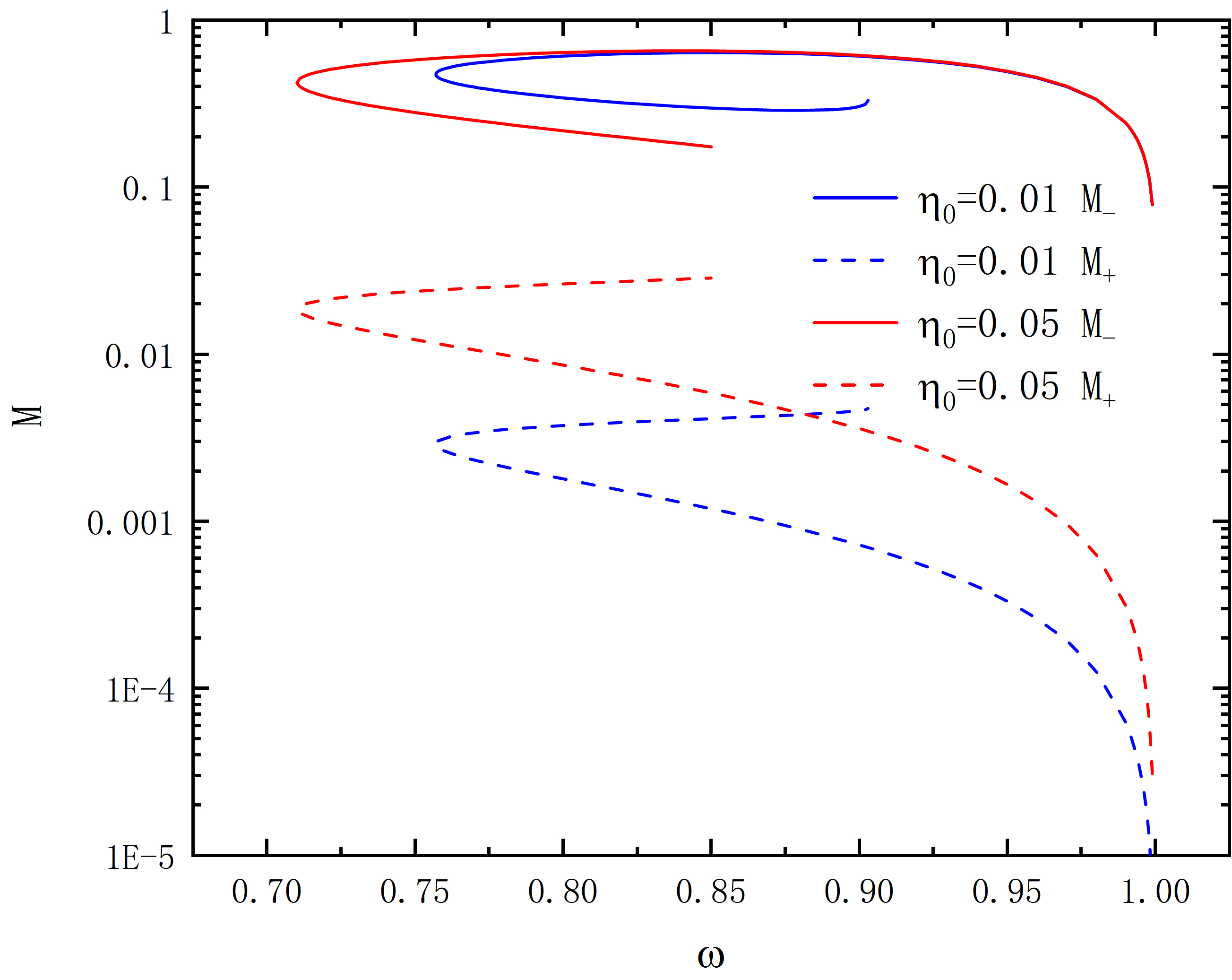}
\includegraphics[height=.26\textheight]{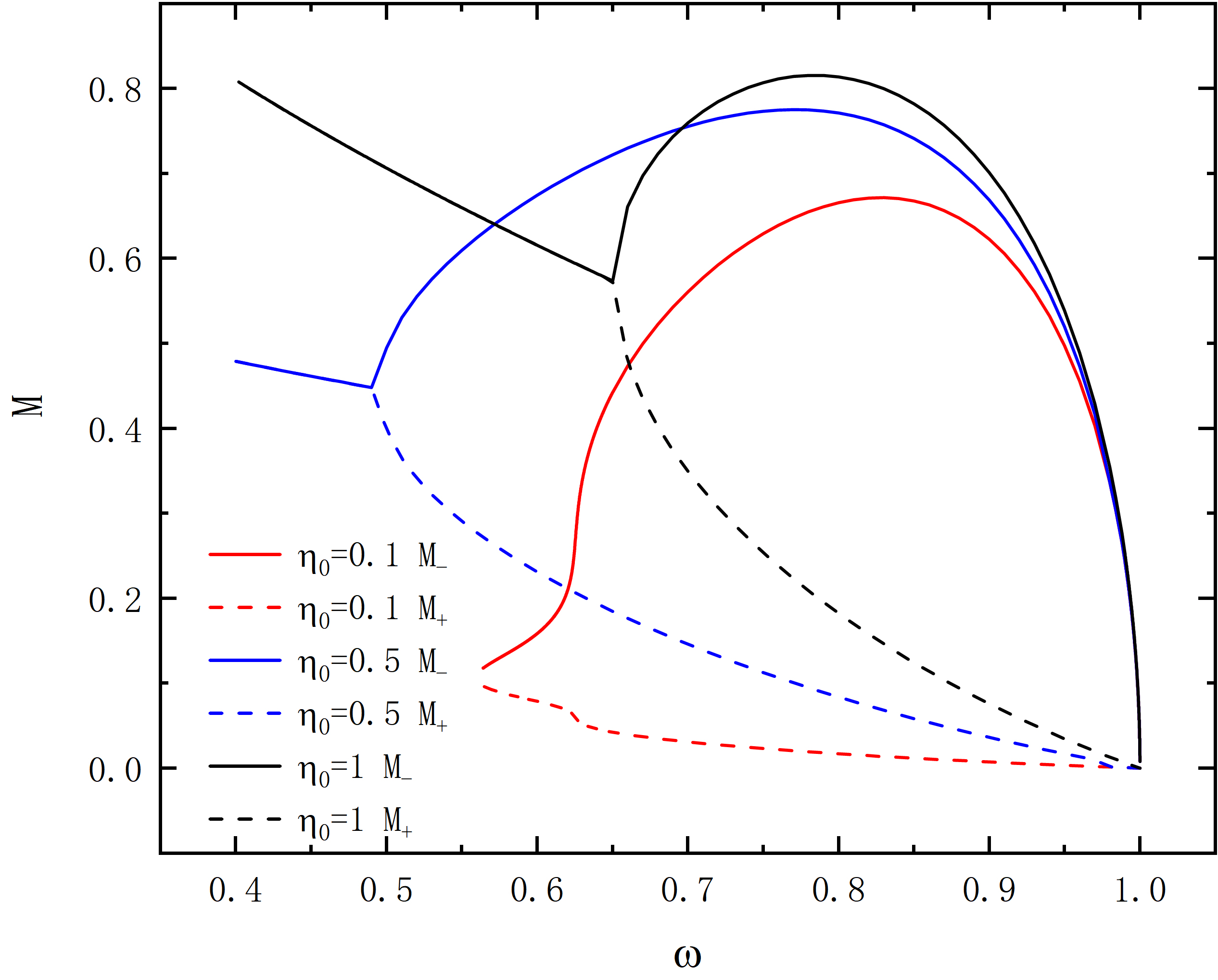}
\includegraphics[height=.26\textheight]{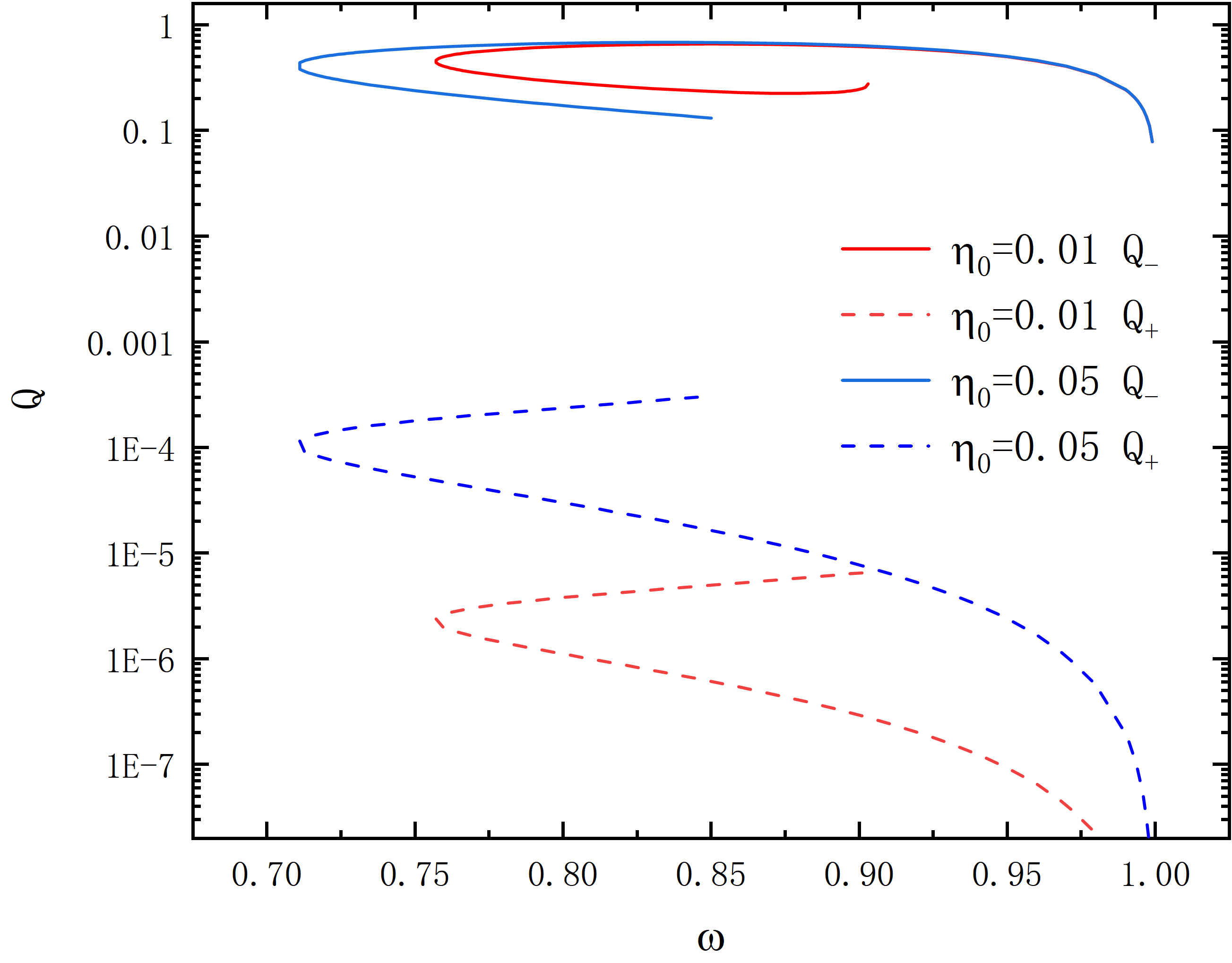}
\includegraphics[height=.26\textheight]{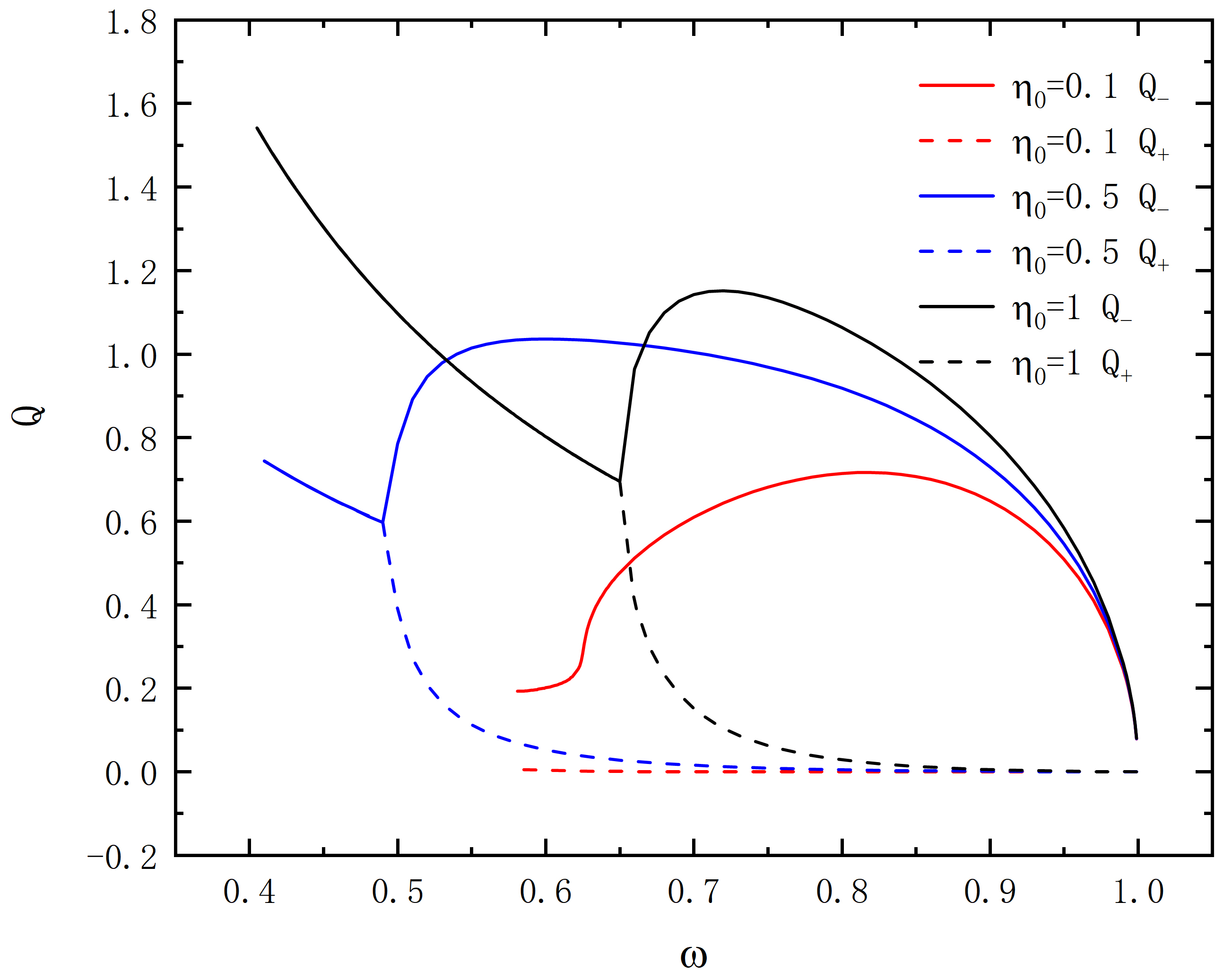}
\end{center}
\caption{The mass $M$ and the charge $Q$ as functions of frequency $\omega$ for $\eta_0=0.01,0.05,0.1,0.5,1$ in the case of asymmetry, all solutions have $c=0$.}
\label{0ASMQ}
\end{figure}

\begin{figure}[!htbp]
\begin{center}
\includegraphics[height=.17\textheight]{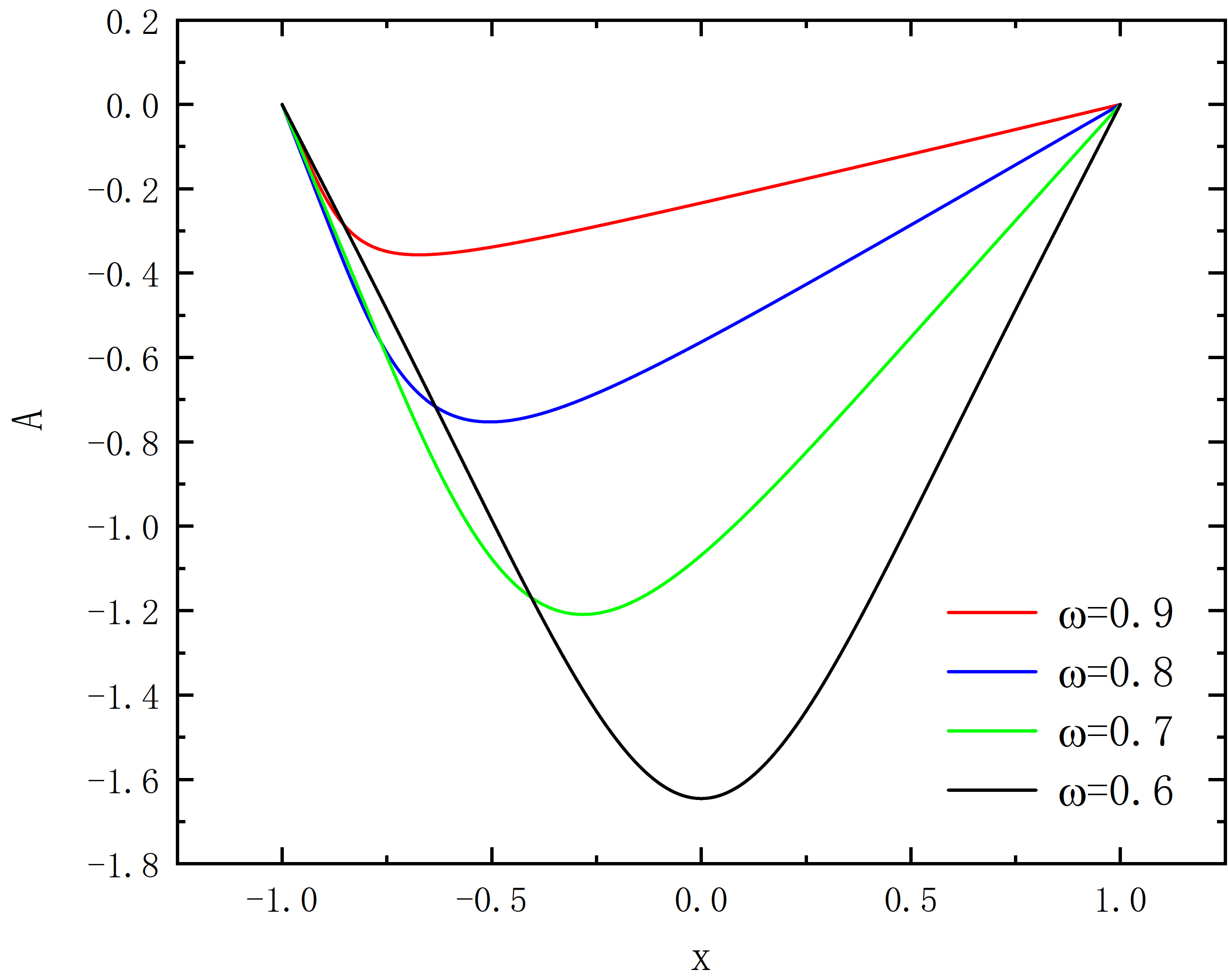}
\includegraphics[height=.17\textheight]{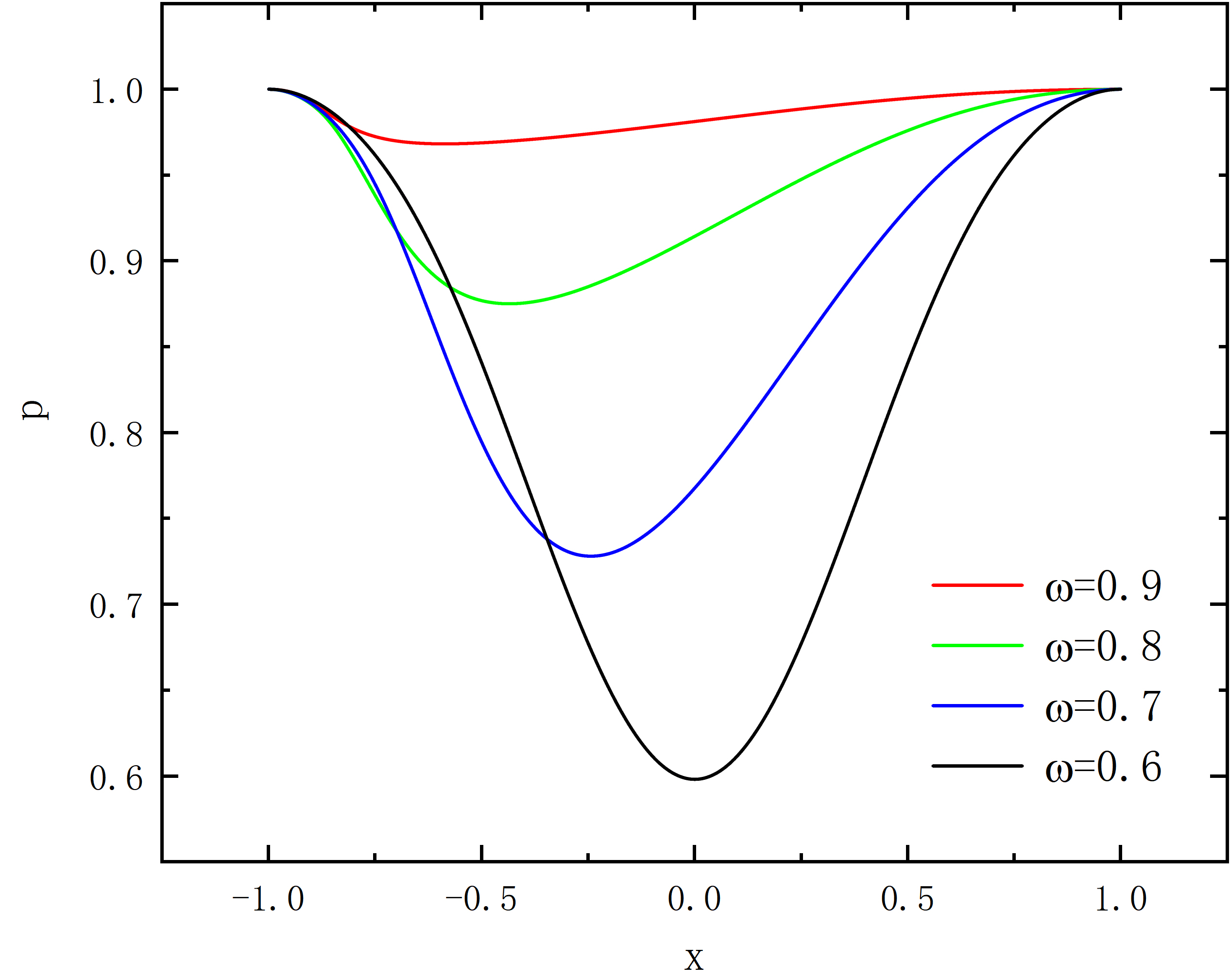}
\includegraphics[height=.17\textheight]{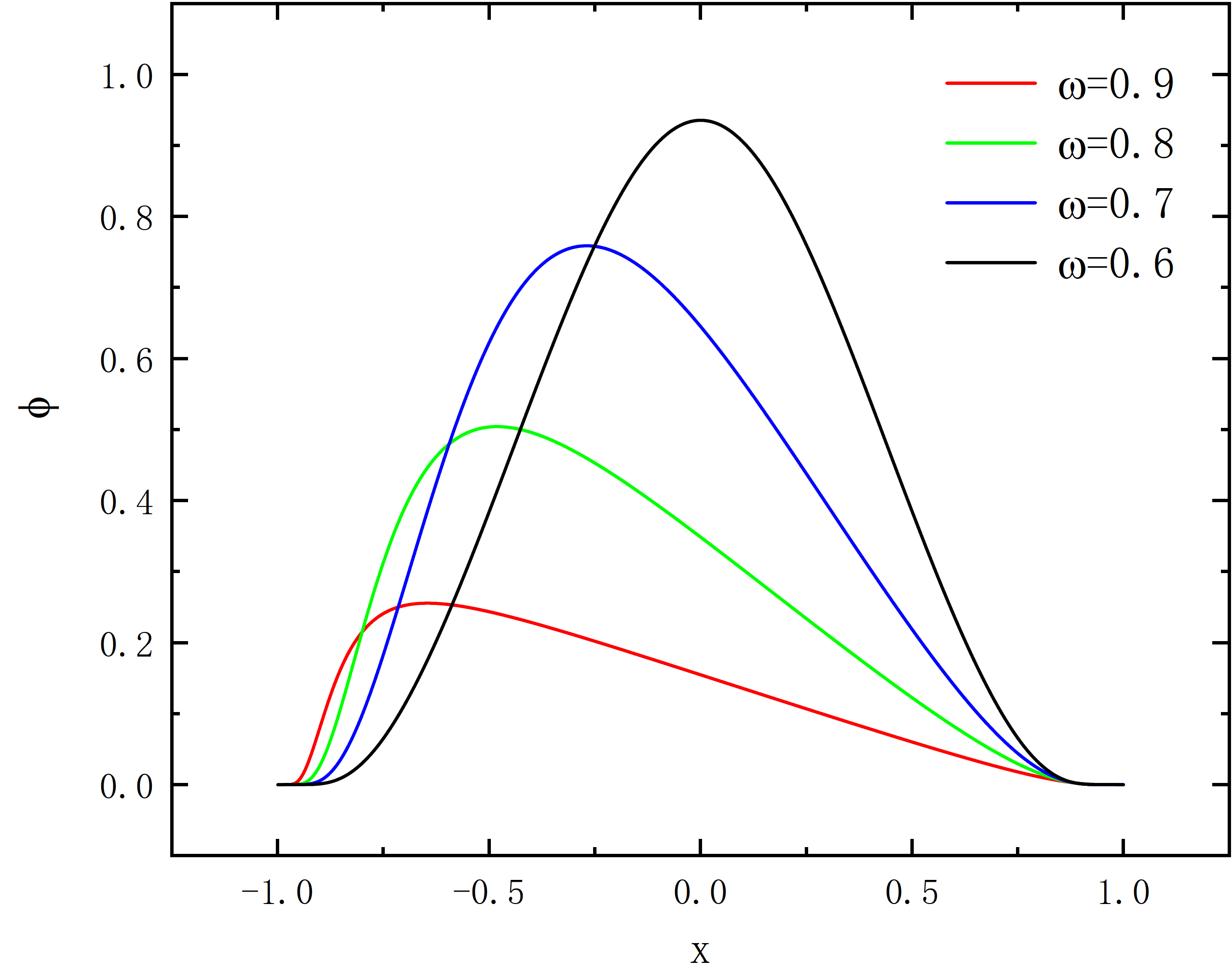}
\end{center}
\caption{Metric function $A$ and $p$ with scalar field $\phi$ as functions of x for $\omega=0.9,0.8,0.7,0.6$ in the case of asymmetry, and $c=0,\eta_0=1$.}
\label{AS1}
\end{figure}

In Fig. \ref{AS_phi_A_p}, we show the results of metric functions $A$, $p$ and field function $\phi$, they also are the solutions in the first branch or the outcome only have one branch. What is like Fig. \ref{Sphi_A_p} is the flat region becomes smaller as the $\eta_0$ increases and. However, the flat region not appear at $x$ near zero but only appear at part of $x\in (-1,0)$, and when $\eta_0$ approach zero, the value of the three functions in the region $x\in (0,1)$ seem to be closed to zero, with the discussion in Fig. \ref{0ASMQ} and Fig. \ref{AS1}, we can find it implies the transform between the symmetry and the asymmetry.

In Fig. \ref{0ASMQ}, we exhibit the mass and charge in the asymmetric solutions at $c=0$, the subscript "$-$" express the integration region is $x\in (-1,0)$ and the region $x\in (0,1)$ is expressed by "$+$", the conclusion is similar to Fig. \ref{0SMQ}, when the $\eta_0$ is small, the solutions are \textit{multi-branch} solutions and only have one branch at the situation of big $\eta_0$.

But there is an interesting phenomenon in $one-branch$ solutions. the curve of $M_+$ and $M_-$ merge together as the parameter $\omega$ becomes small, so are $Q_+$ and $Q_-$. In Fig. \ref{AS1}, we present the evolution of functions at $\eta_0=1$, the graph clearly displays the reason of this phenomenon. When $\omega$ goes to small, the asymmetric solutions turn into symmetric solutions. Such outcome declares that in some setting of parameters, the asymmetric solutions are non-existent. These results indicate that the transformation from asymmetry to symmetry can be achieved.

\begin{figure}[!htbp]
\begin{center}
\includegraphics[height=.28\textheight]{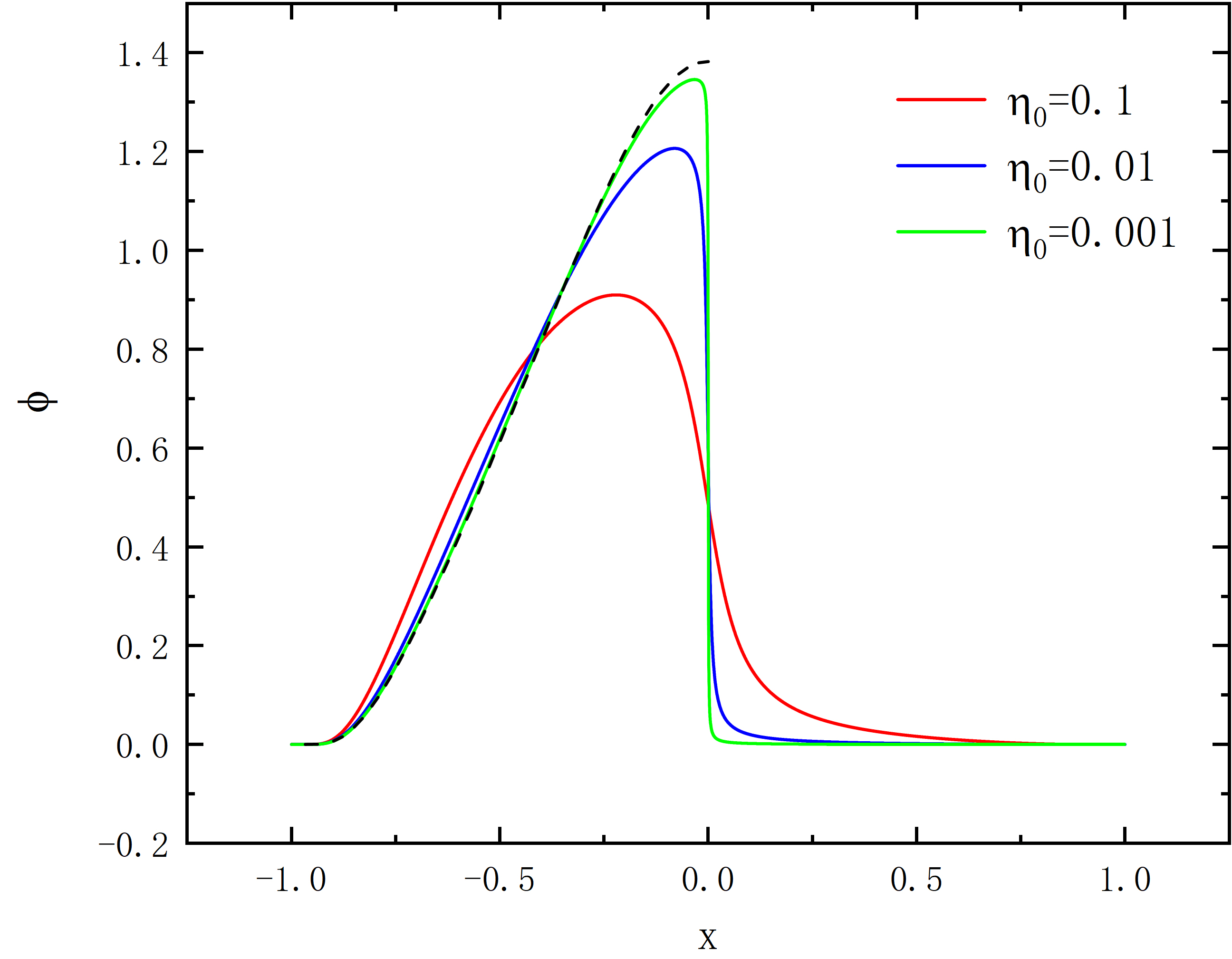}
\end{center}
\caption{The scalar field function $\phi$ as function of x for $\eta_0=0.1,0.01,0.001$ in the case of asymmetry, the black line is the symmetric solution at $\eta_0=0$, all solutions have $c=0,\omega=0.77$.}
\label{AS0}
\end{figure}

Also interesting are the solutions of $\eta_0 \rightarrow 0$, In Fig .\ref{0ASMQ}, we can discover $M_+$ and $Q_+$ have the trend at $\eta_0=0.01,0.05$, what's more, the curve shape of $M_-$ and $Q_-$ are similar to the symmetric solutions. Fig. \ref{AS0} exhibit the trend of $\phi$ as $\eta_0$ approaches zero in the first branch. The figure expresses that the asymmetric solution turns into the the single boson star solution without wormhole in region $x\in (-1,0)$ and magnitude approaches zero in region $x\in (0,1)$.

 The transition of asymmetrical solutions can be concluded in follows: when the value of $\eta_0$ is high, there is a reduction in asymmetry and an increase in symmetry along with the decrease of $\omega$. In the solution of $\eta_0 \rightarrow 0$, the transformation is incomplete, the magnitude is constrained in the region $x \in (-1,0)$, in other words, the outcome degenerates to a single scalar field solution in one asymptotically flat region.

\begin{figure}[!htbp]
\begin{center}
\includegraphics[height=.26\textheight]{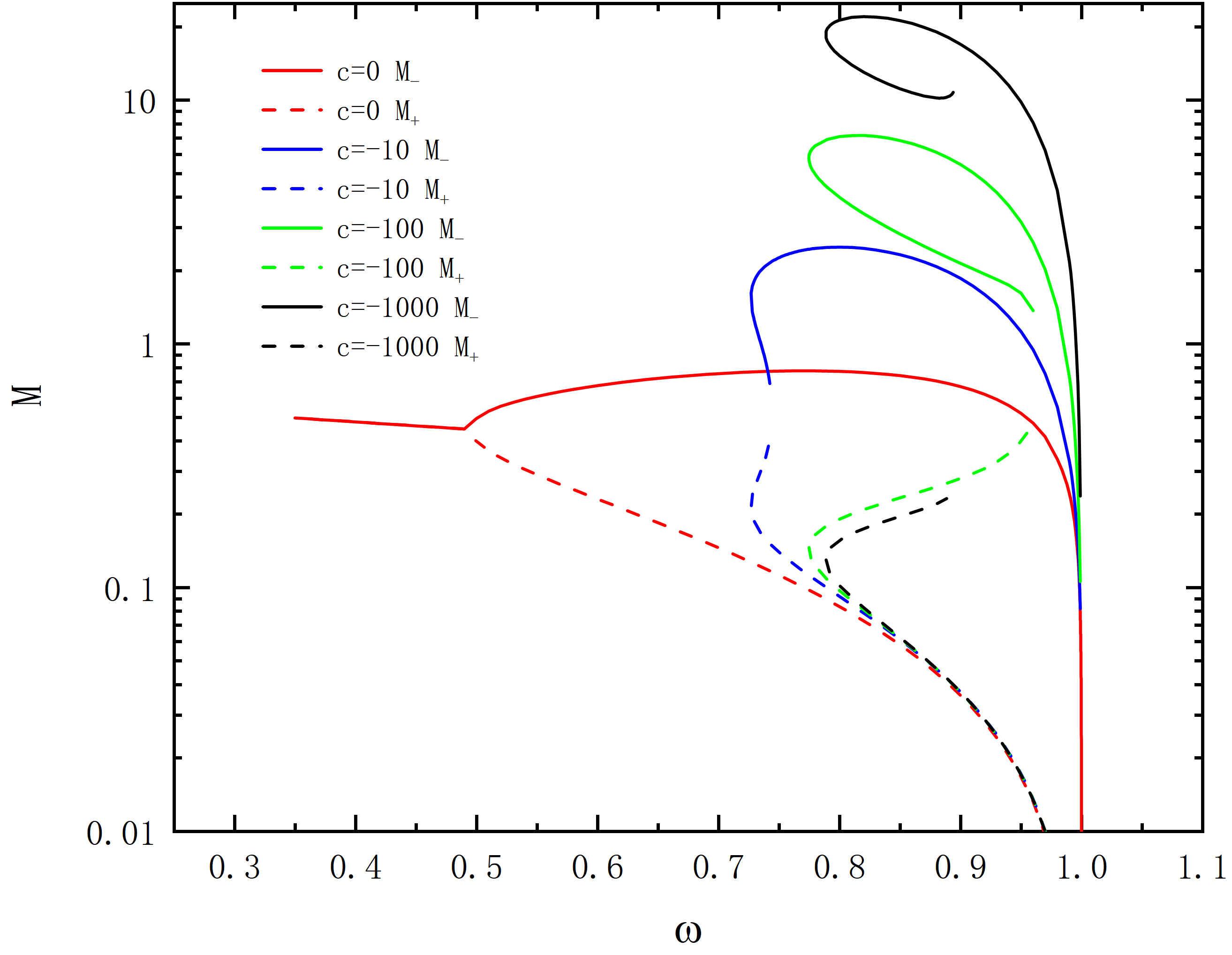}
\includegraphics[height=.26\textheight]{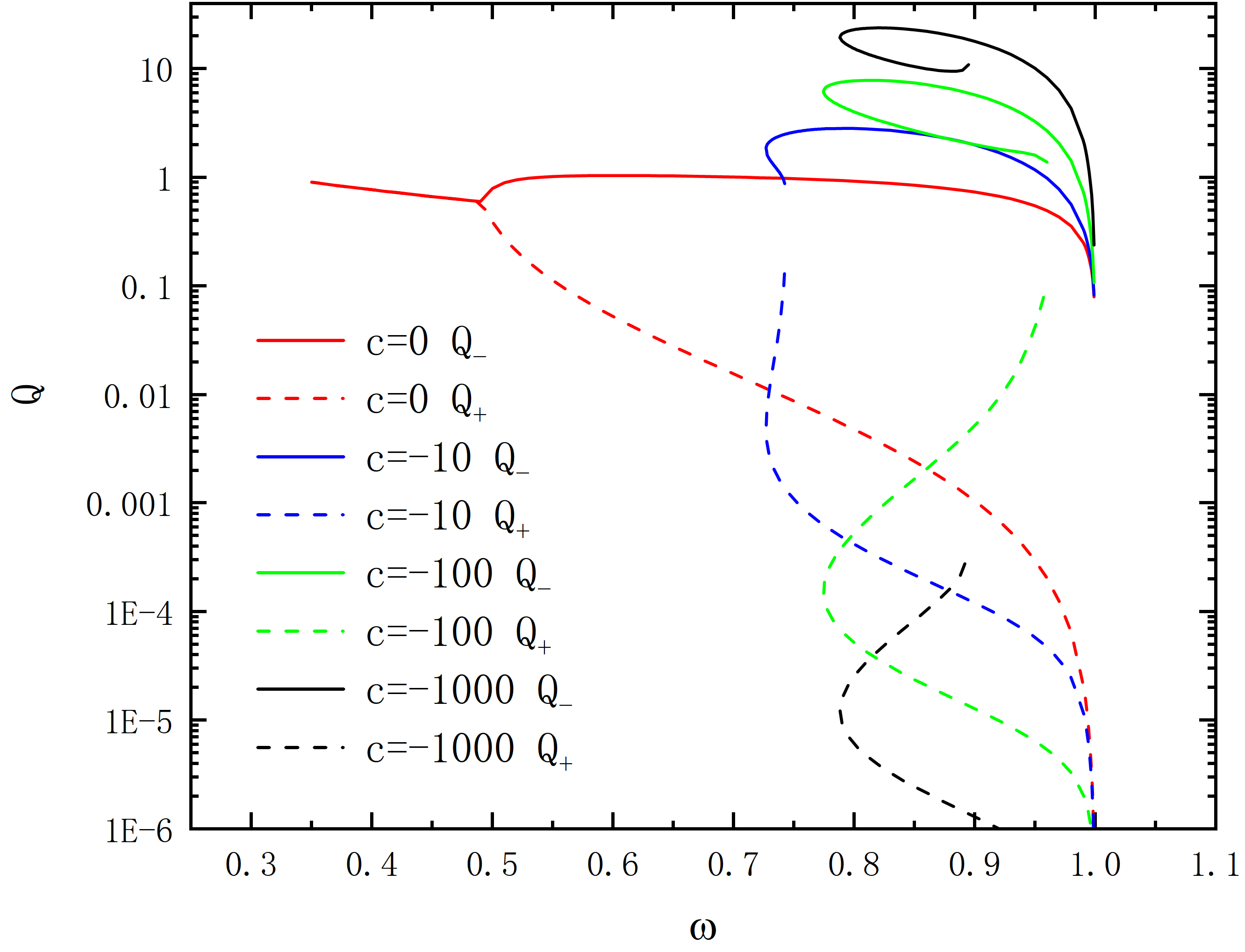}
\end{center}
\caption{The mass $M$ and the charge $Q$ as functions of frequency $\omega$ for $c=0,-10,-100,-1000$ in the case of symmetry, all solutions have $\eta_0=0.5$.}
\label{CASMQ}
\end{figure}

In Fig. \ref{CASMQ}, we unveil the value of $M$ and $Q$ where $c \neq 0$ and $\eta_0=0.5$, as excepted, the conclusion is homologous to the symmetric solutions, the shape of curve is spiral line when absolute value of $c$ is large, corresponds to the outcome at low value of $\eta_0$ at $c=0$, and the results
in two asymptotically flat regions convergence, represent the solutions have become the symmetric solutions.

\subsection{Wormhole geometries}

Next, we intend to explore the geometry of wormhole. Firstly, we fix $t$ and $\theta$ ($\theta = \pi /2$) to obtain the equatorial plane, it's useful for observe the shape of wormhole and count the number of throats. Embedding the equatorial plane in Euclidean space:

\begin{figure}[!htbp]
\begin{center}
\includegraphics[height=.22\textheight]{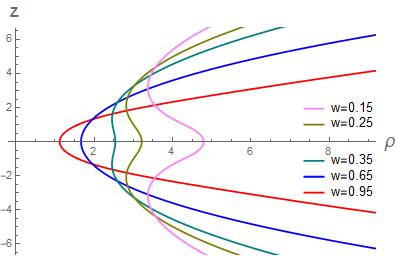}
\includegraphics[height=.22\textheight]{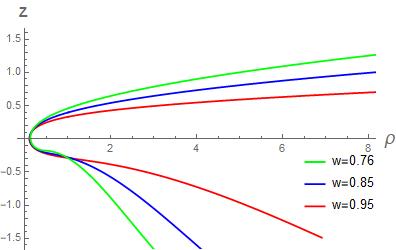}
\end{center}
\caption{Geometry of throats: isometric embedding in two dimensional view. Symmetric solution for $\eta_0=1$(left panel), asymmetric solution for $\eta_0=0.1$(right panel), all solutions have $c=0$.}
\label{EMB}
\end{figure}

\begin{equation}
\begin{aligned}
ds^2 &= p e^{-A}  d \eta^2 + p e^{-A} h   d \phi^2 \, \\
&= d \rho^2 + dz^2 + \rho^2 d \phi^2   \,.
\end{aligned}
\end{equation}
here we choose the cylindrical coordinates ($\rho,\varphi,z$) to express it, and the representation of $\rho$ and $z$ can be calculated:

\begin{equation} \label{formula_embedding}
 \rho(\eta)= \sqrt{ p(\eta) e^{-A(\eta)} h(\eta) } ,\;\;\;\;\;\;\;\;\;\;   z(\eta) =  \int  \sqrt{ p(\eta) e^{-A(\eta)}  -   \left( \frac{d \rho}{d \eta} \right)^2    }     d \eta .
\end{equation}

Coordinate $\rho$ can be seen as the circumferential radius $R_c$, and it should be the minimum value in the case of throat, so it sates:

\begin{equation}
 \frac{d R_c}{d r} \Bigg|_{\eta=\eta_{\text{th}}} = 0\,, \qquad \frac{d^2 R_c}{d r^2}  \Bigg|_{\eta=\eta_{\text{th}}} > 0 \,,
\end{equation}
$\eta_{th}$ corresponds to the radius of throat.

\begin{figure}[!htbp]
\begin{center}
\includegraphics[height=.23\textheight]{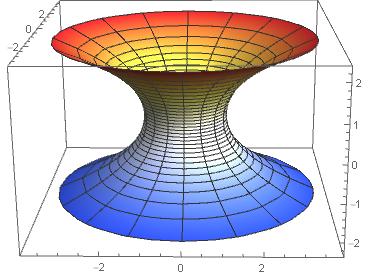}
\includegraphics[height=.23\textheight]{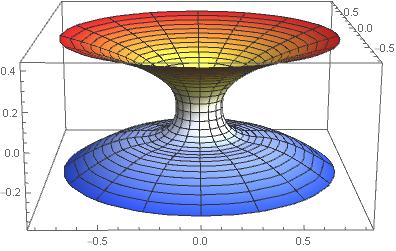}
\end{center}
\caption{Geometry of single throat: isometric embedding in three dimensional view. Symmetric solution for $\omega =0.95, \eta_0=1$ (left panel), asymmetric solution for $\omega =0.76, r_0=0.1$(right panel), all solutions have $c=0$.}
\label{Geo1}
\end{figure}

\begin{figure}[!htbp]
\begin{center}
\includegraphics[height=.30\textheight]{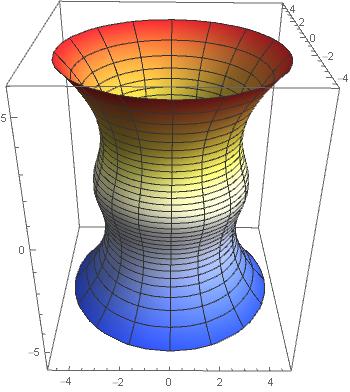}
\end{center}
\caption{Geometry of double throat: isometric embedding of symmetric solution with parameters for $\omega =0.25, \eta_0=1$.}
\label{Geo2}
\end{figure}

In Fig. \ref{EMB}, we exhibit the equatorial plane which embedding in Euclidean space both in symmetry and asymmetry. For the symmetric solutions($\eta_0$=1), we are informed of the wormhole has just one throat when the $\omega$ is high and the trend of having double throats has appeared near $\omega=0.35$. For the asymmetrical solutions($\eta_0=0.1$), there is only the trend of becoming more asymmetric, not occurrence more throats. Besides, we showcase the geometric properties of throats in three dimensional view in Fig. \ref{Geo1} and Fig. \ref{Geo2} and we haven't find more throats at other setting of parameters.

\section{Excited state}\label{sec5}
Following the study about ground state in Sec~.\ref{sec4}, we make a study of excited state in this section. The difference between the ground state and the excited state can be shown in the curve of $\phi$, if there are some zero points in region $(-1,1)$, the solutions are excited state. Based on the number of zero points which we use integer $n$ to represent, the situations are further distinguished, for the solutions of asymmetry, the number $n$ also represent the level of excited state, for example, $n=1$ on behalf of the first excited state. However, due to symmetry, $n$ must be even number, and here $\frac{n}{2}$ stands for the level of excited state.

We solve the situation of first excited state and second excited state both in symmetry and asymmetry. And we use same algorithm and coordinate regulations in Sec~.\ref{sec4} to calculate. What's more $c=0, \eta_0=0.5$ is applied to this section.

\subsection{Numerical results}

\begin{figure}[!htbp]
\begin{center}
\includegraphics[height=.17\textheight]{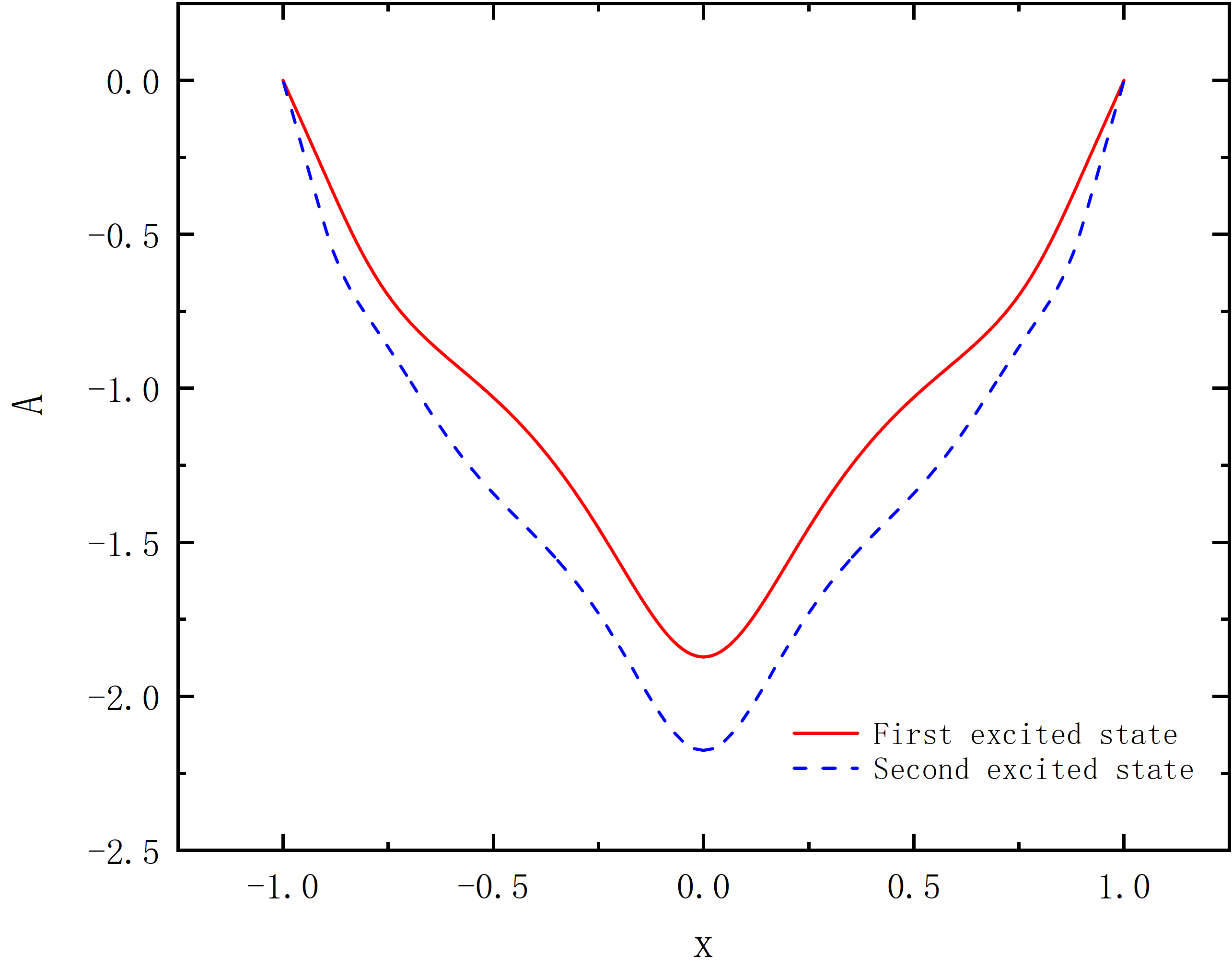}
\includegraphics[height=.17\textheight]{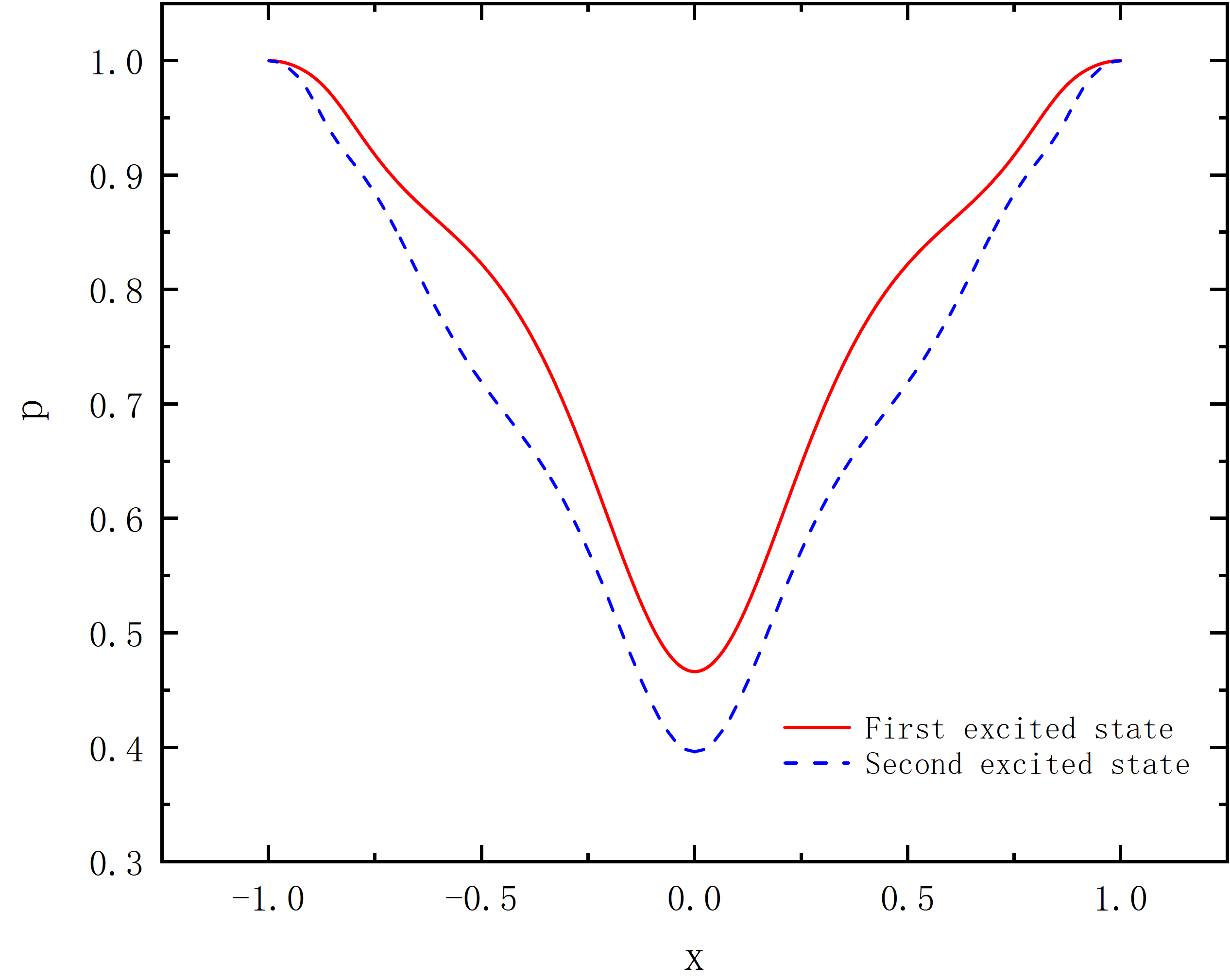}
\includegraphics[height=.17\textheight]{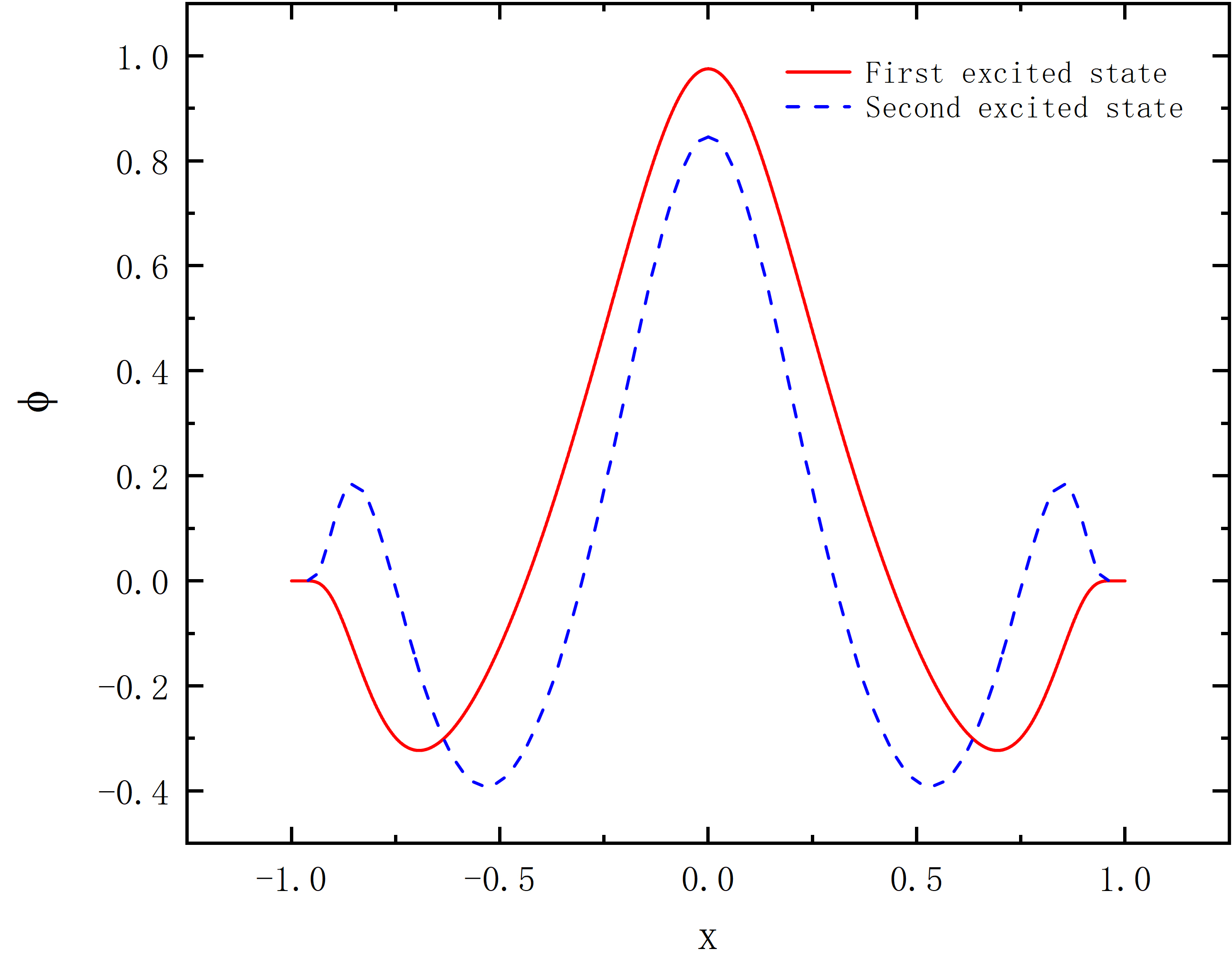}
\includegraphics[height=.17\textheight]{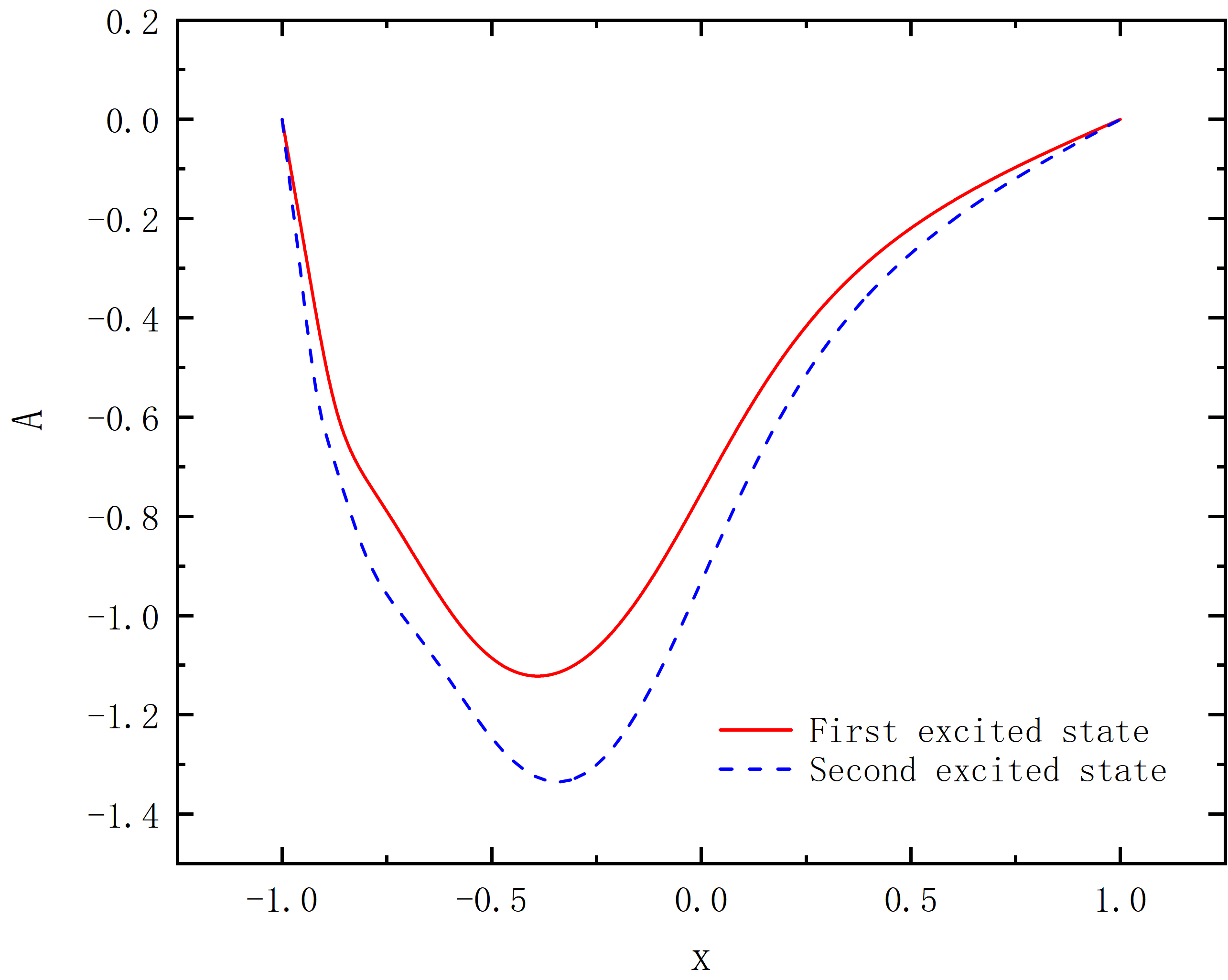}
\includegraphics[height=.17\textheight]{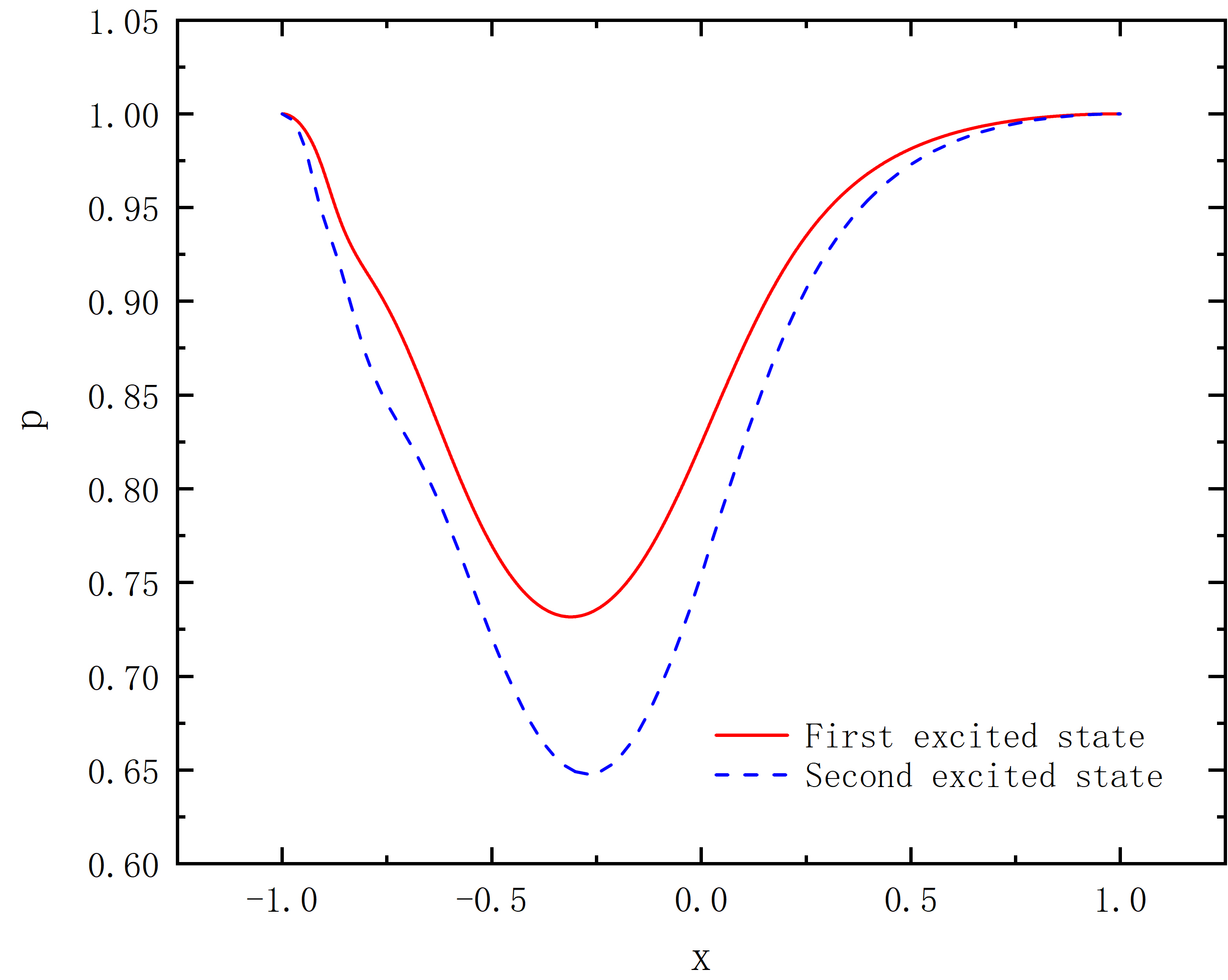}
\includegraphics[height=.17\textheight]{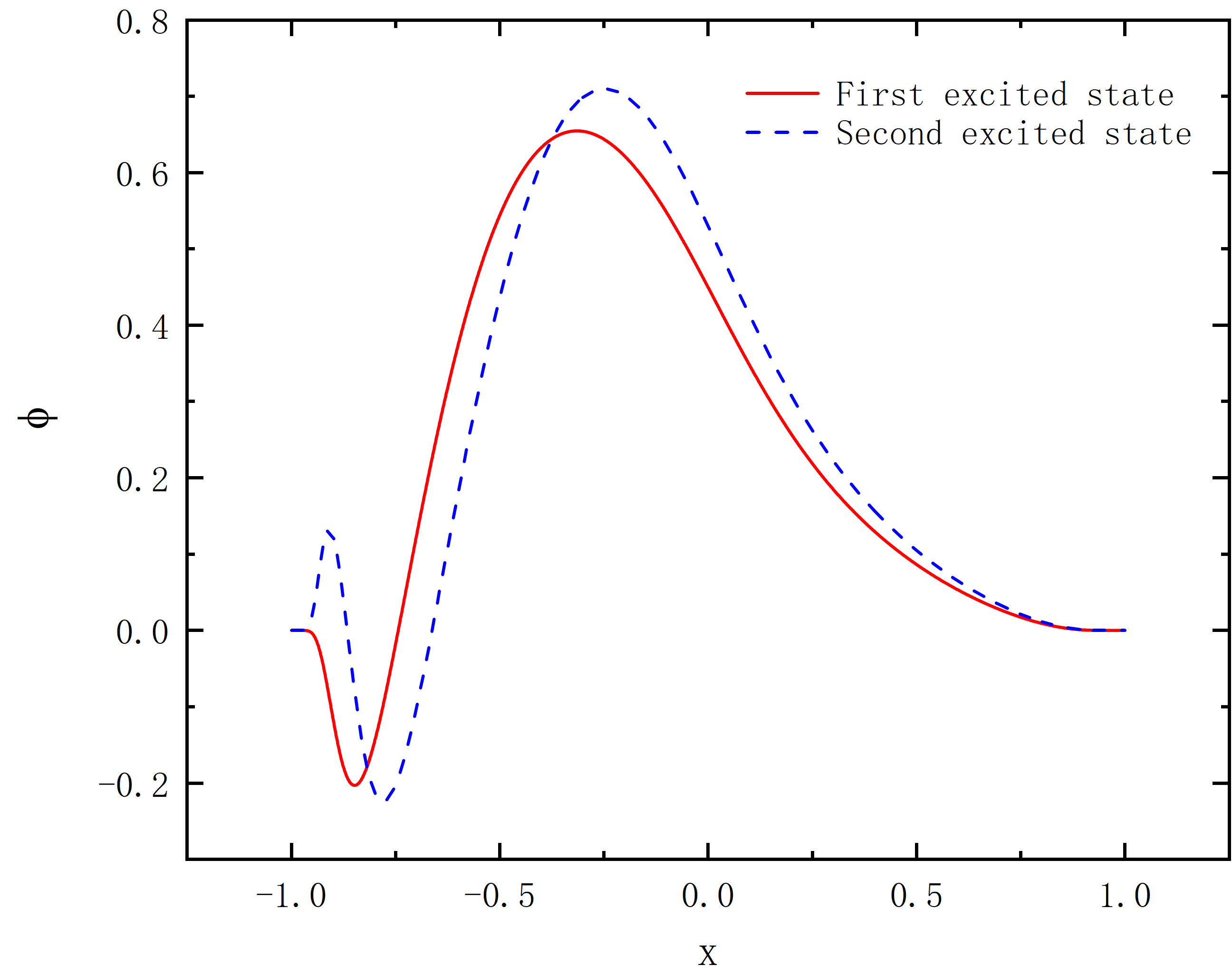}
\end{center}
\caption{Metric function $A$ (left two panels) and $p$ (middle two panels) with scalar field $\phi$ (right two panels) as functions of x for $\eta_0=0.5$ in the case of symmetry and asymmetry, and $\omega=0.8$.}
\label{asands}
\end{figure}

In Fig.\ref{asands}, we exhibit the curve of metric functions and field function both in symmetry and asymmetry and if there are $multi-branch$ solutions, we choose the first branch to display. The results reveal the minimum value of $A$ and $p$ become smaller with the increase of excited state level. However, $\phi$ isn't similar to metric functions, it's maximum value performs different trend in symmetric solutions and asymmetric solutions. We haven't find zero points appear in region $(0,1)$ for asymmetric solutions.

\begin{figure}[!htbp]
\begin{center}
\includegraphics[height=.16\textheight]{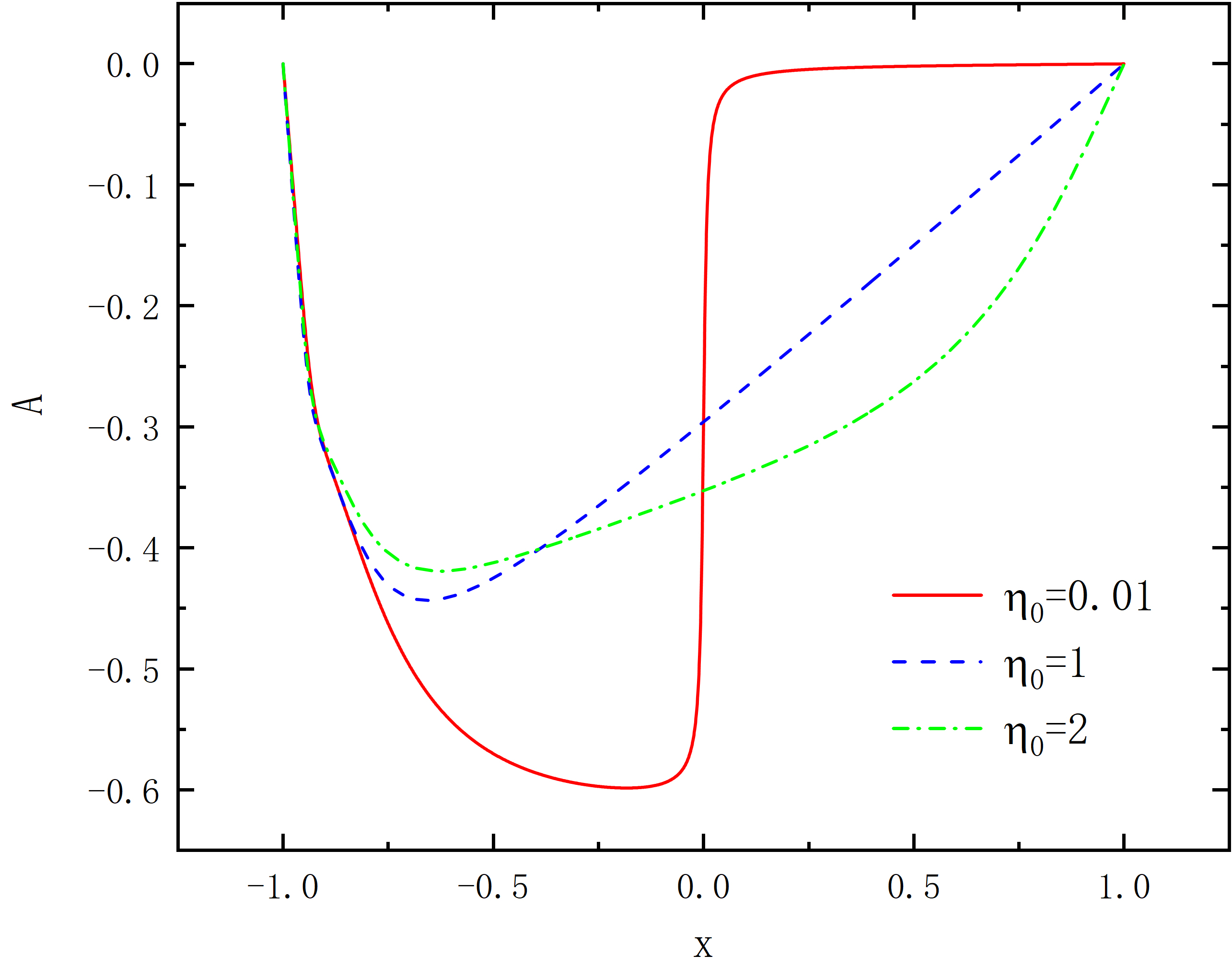}
\includegraphics[height=.16\textheight]{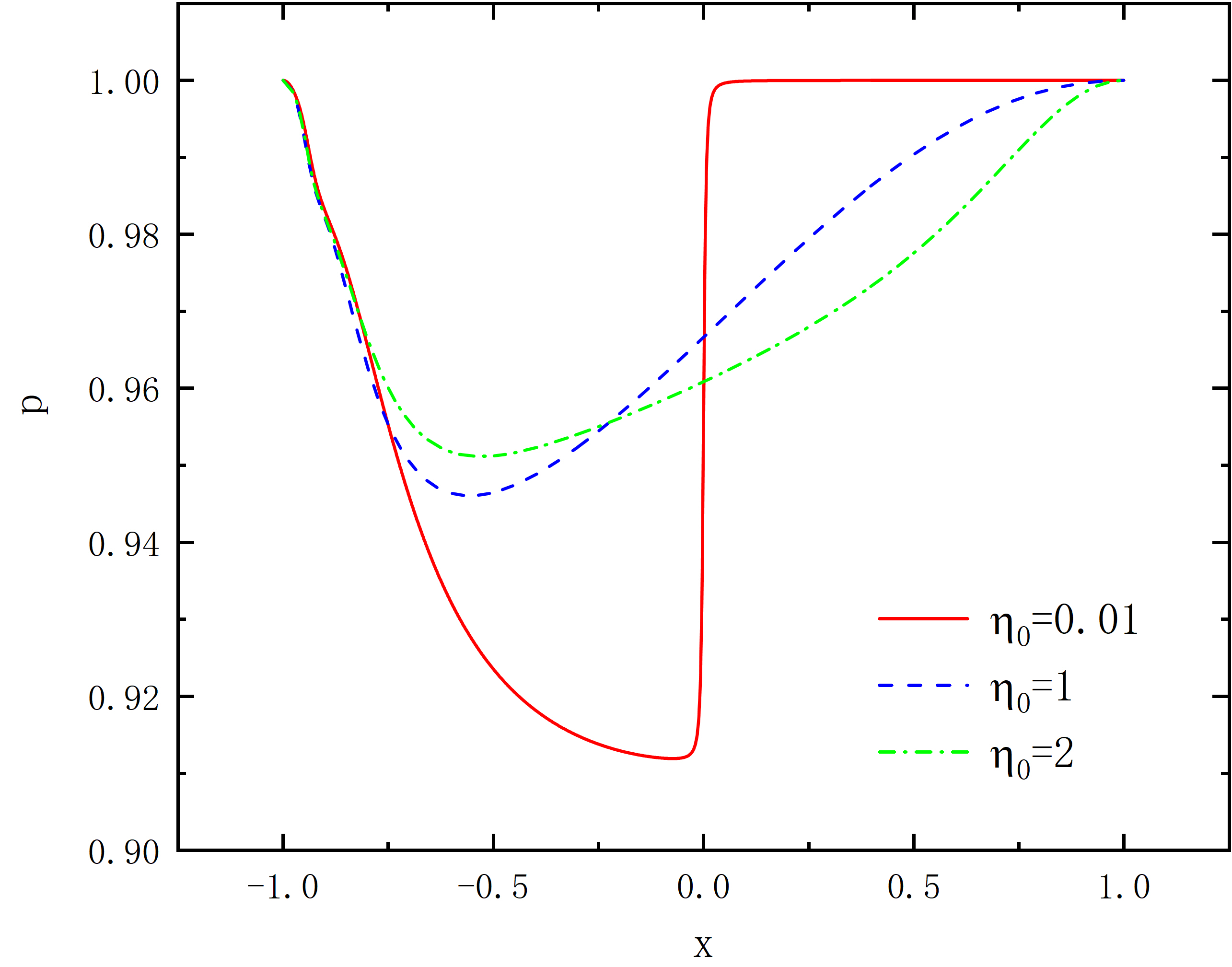}
\includegraphics[height=.16\textheight]{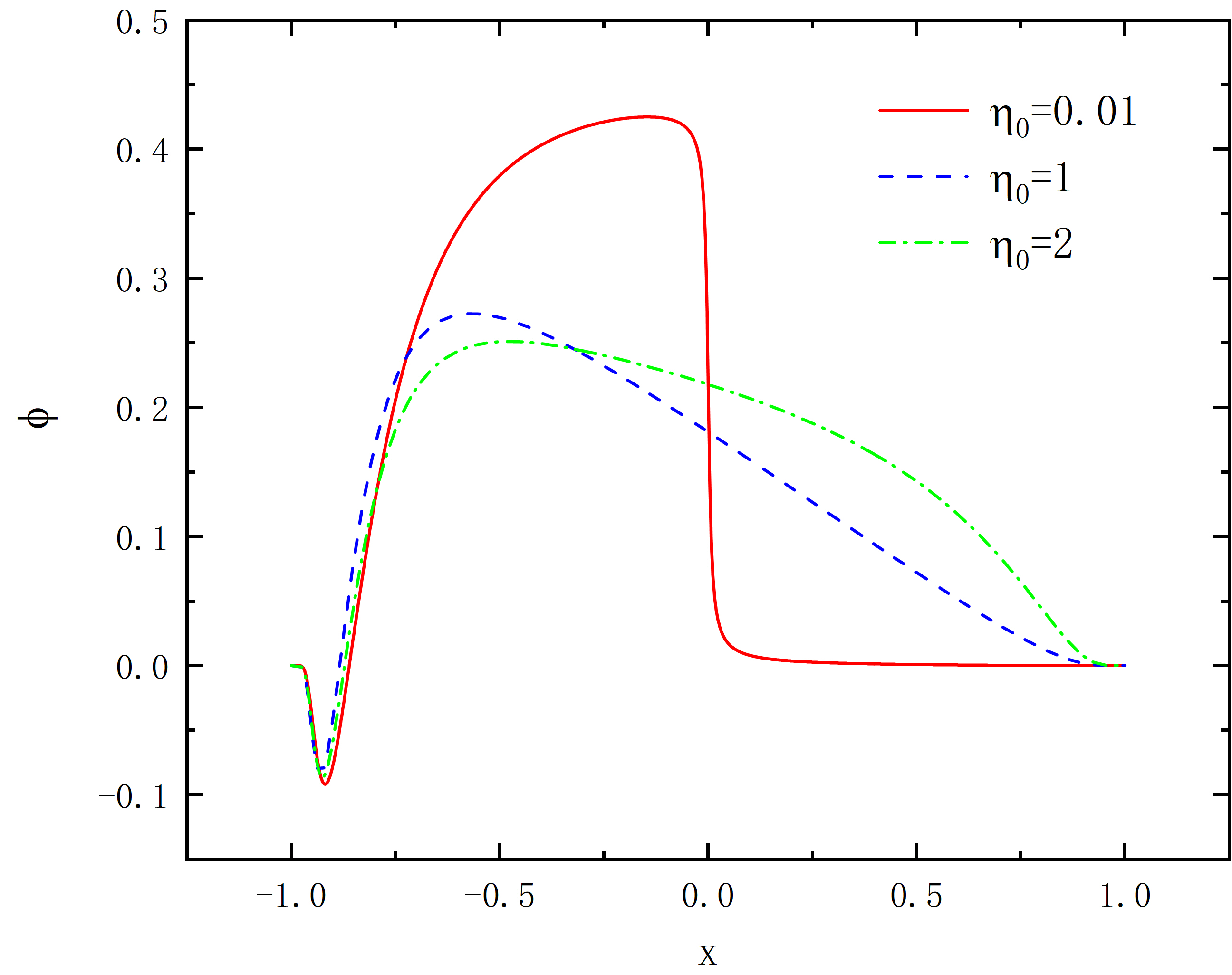}
\includegraphics[height=.16\textheight]{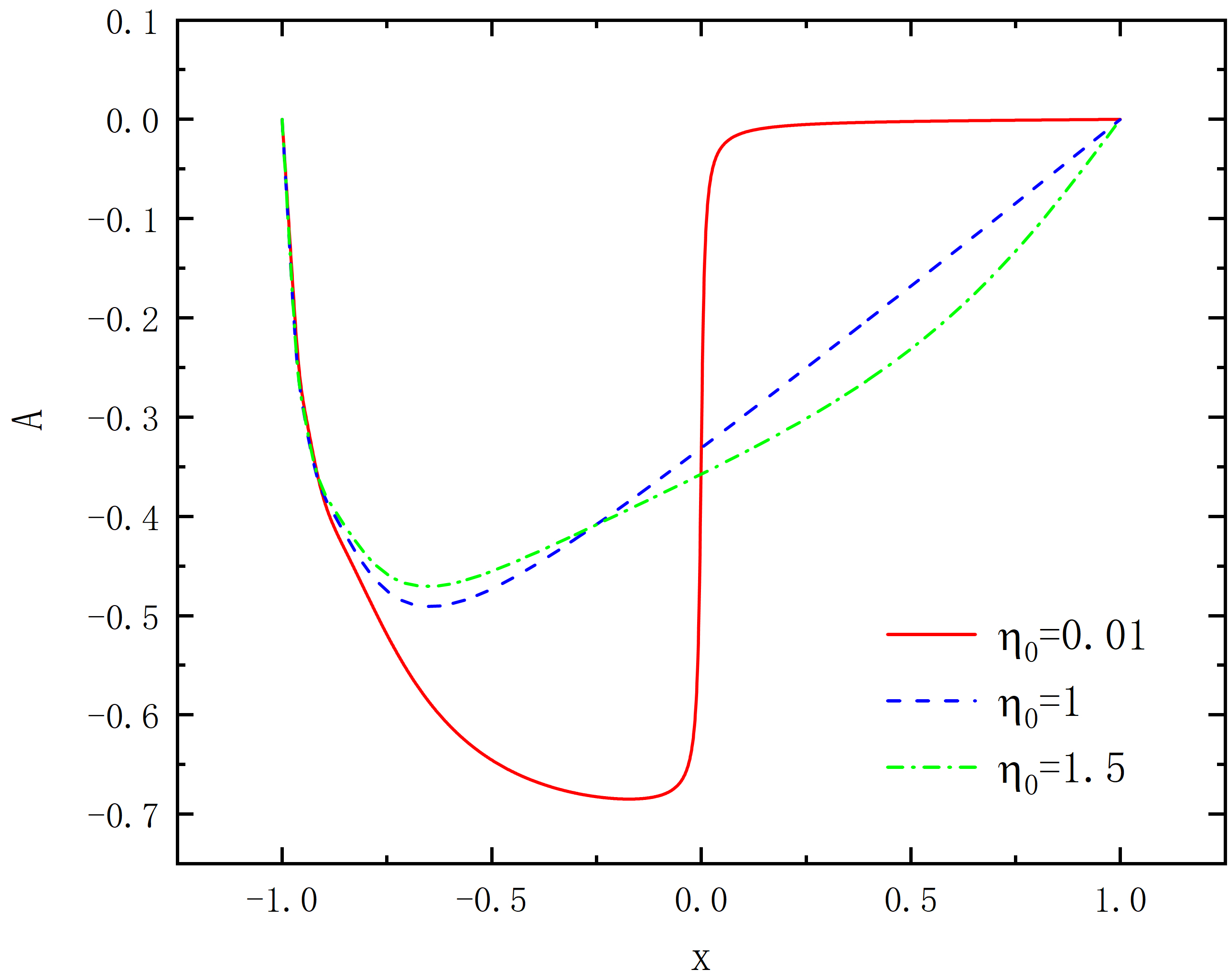}
\includegraphics[height=.16\textheight]{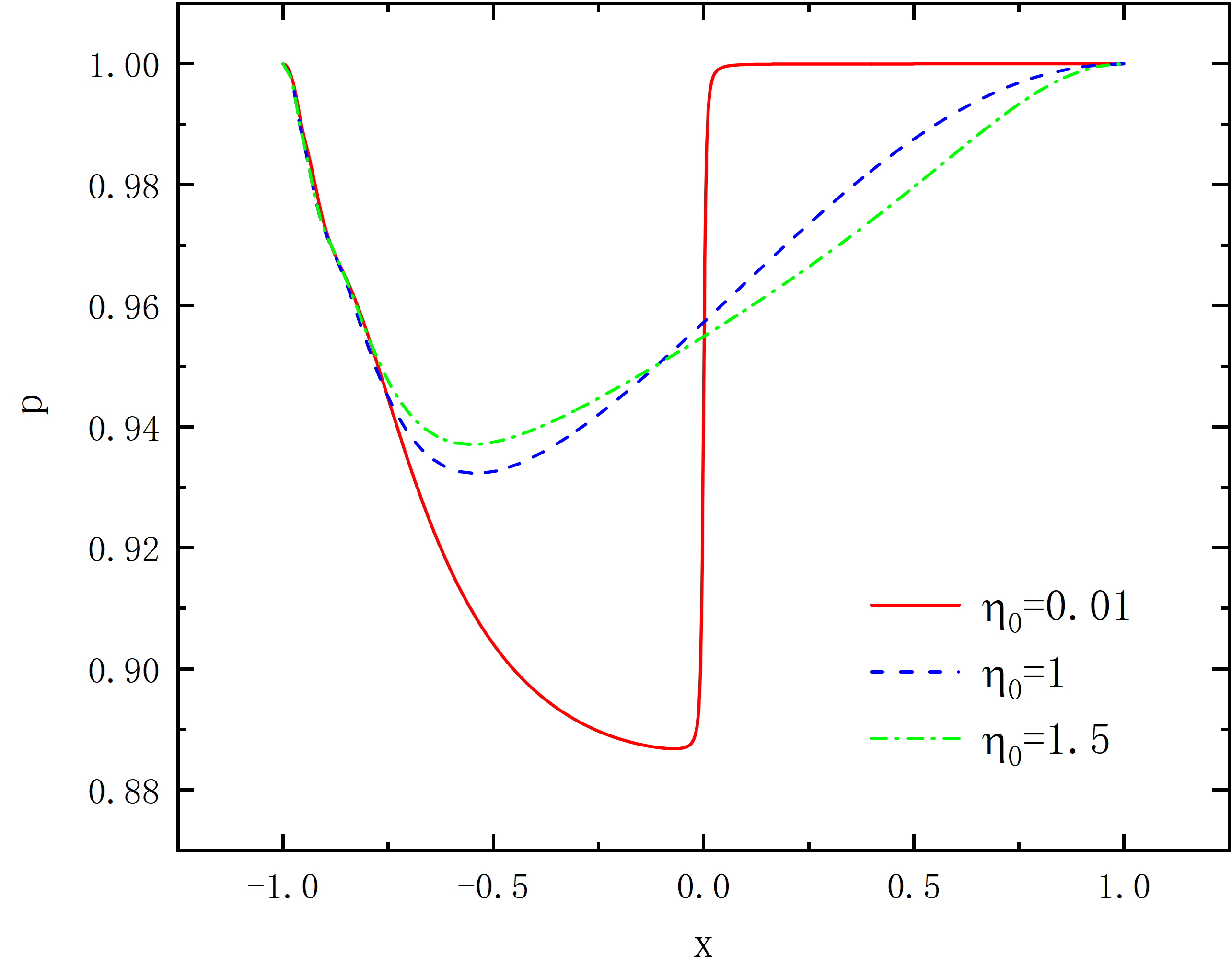}
\includegraphics[height=.16\textheight]{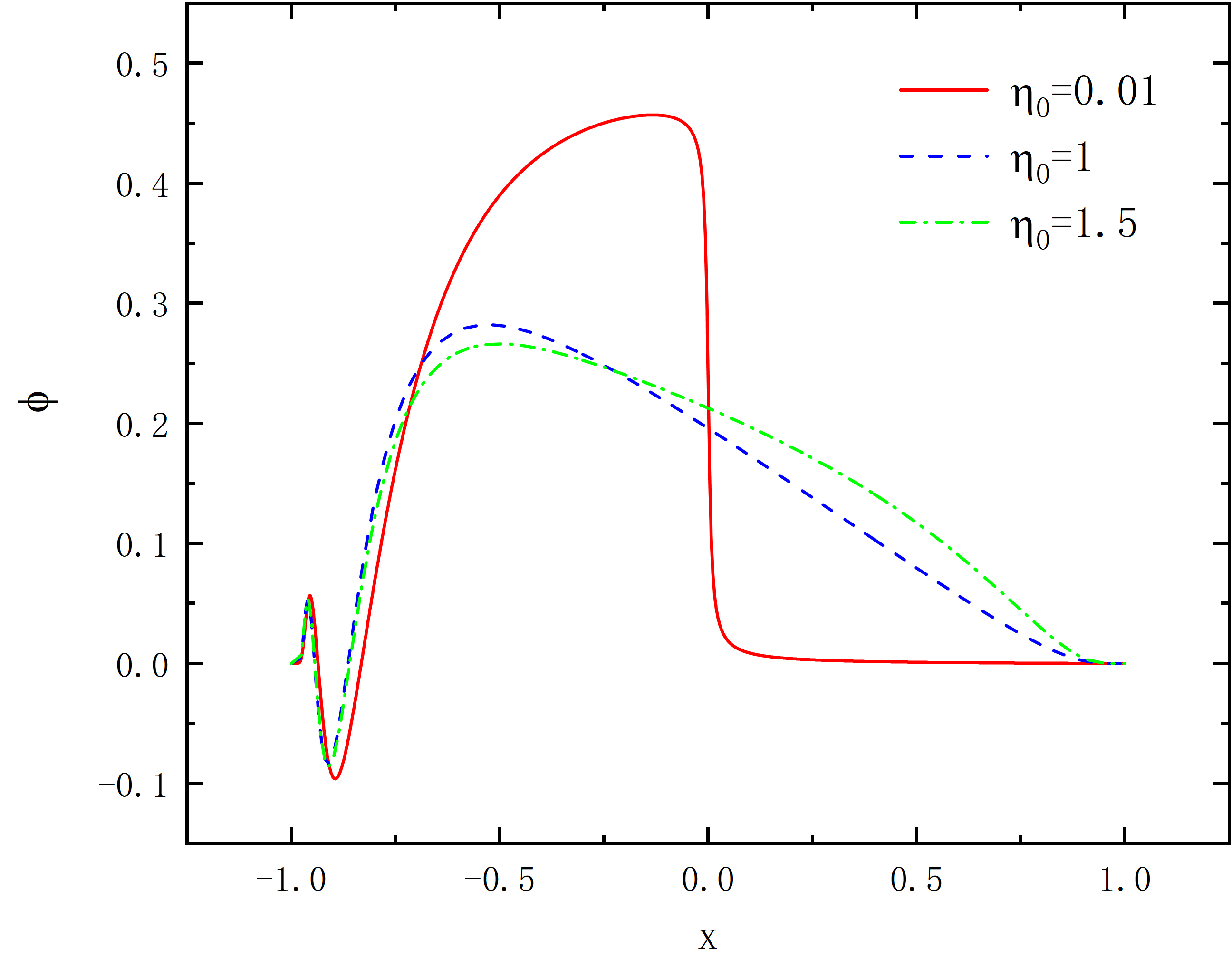}
\end{center}
\caption{Metric function $A$ (left two panels) and $p$ (middle two panels) with scalar field $\phi$ (right two panels) as functions of x for $\eta_0=0.01,1,1.5$ in the case of symmetry and asymmetry, and $\omega=0.9$.}
\label{asandsr}
\end{figure}

\begin{figure}[!htbp]
\begin{center}
\includegraphics[height=.22\textheight]{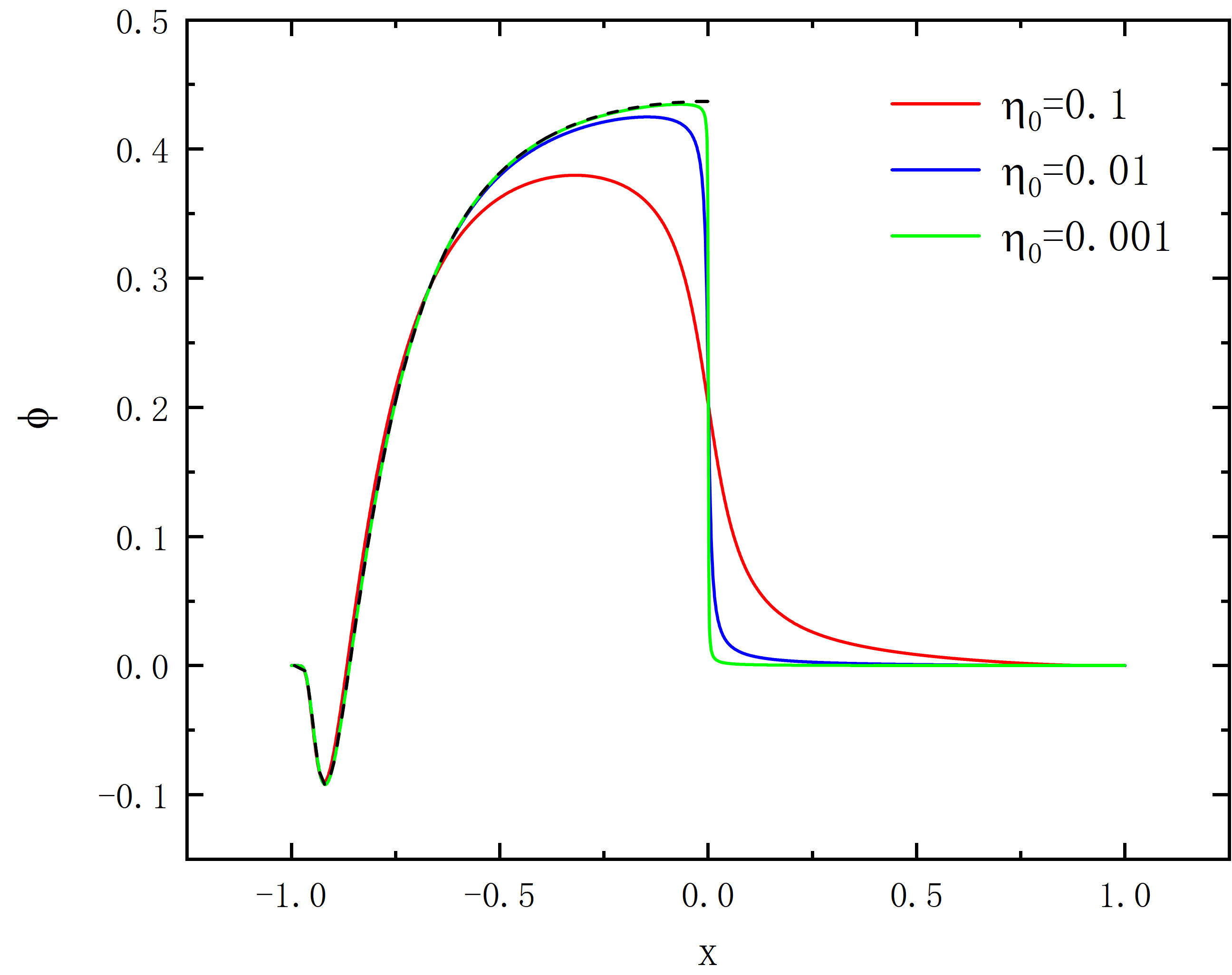}
\includegraphics[height=.22\textheight]{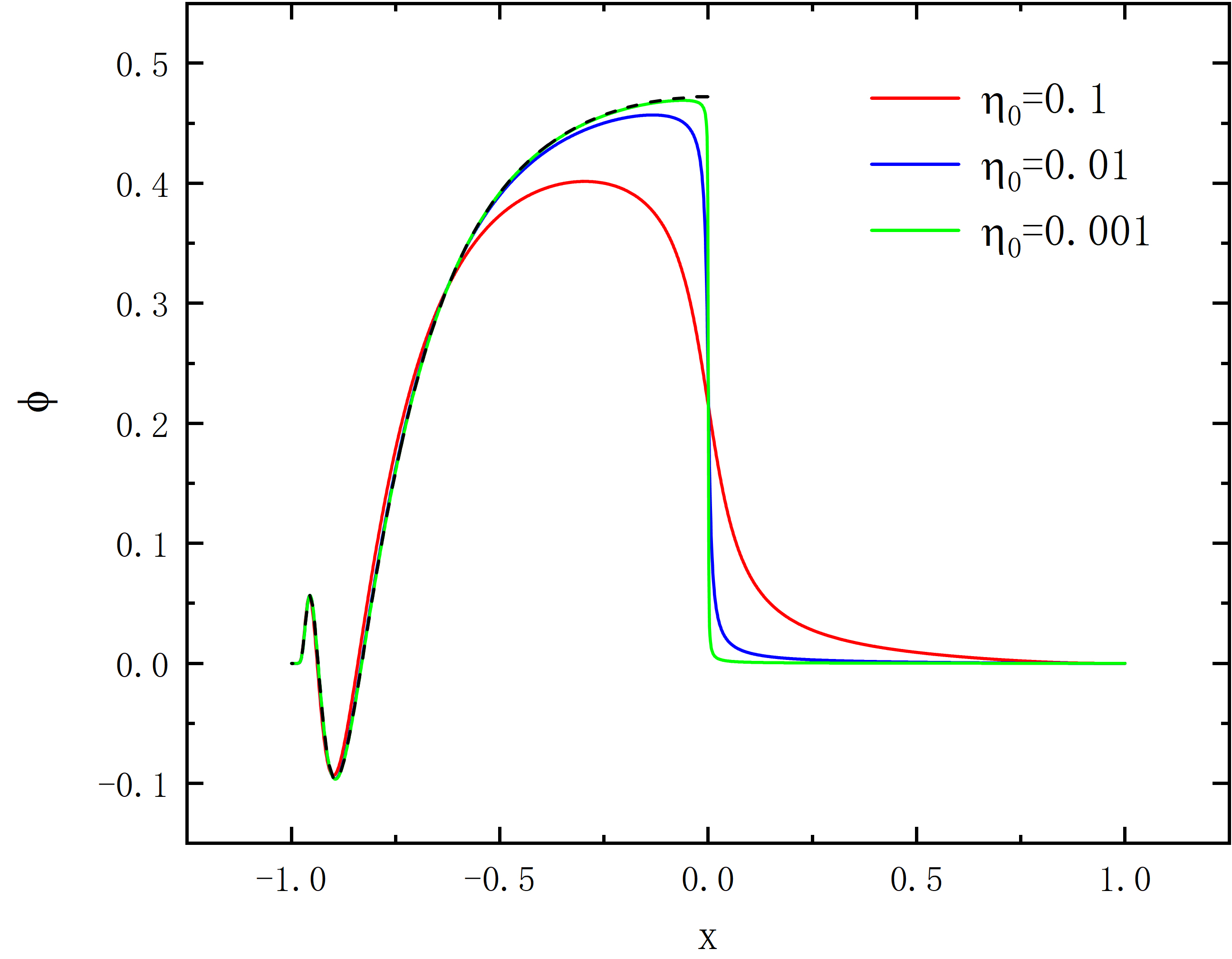}
\end{center}
\caption{The scalar field function $\phi$ as function of x for $\eta_0=0.1,0.01,0.001$ for asymmetric solutions about first excited state and second excited state, the black line is the symmetric solution at $\eta_0=0$, all solutions have $\omega=0.9$.}
\label{AS012}
\end{figure}

Fig.\ref{asandsr} shows the difference in asymmetric solutions with the variation of $\eta_0$, the shift is analogous to situation of ground state, the reduction of $\eta_0$ leads to an increase in the maximum absolute value of these three functions. The depicted trend in Fig.\ref{AS012} also shares similarities with that of the ground state, as $\eta_0$ decreases to 0, all solutions converge into a single boson star solution without a wormhole in the range $x \in (-1,0)$, and the magnitude approaches zero in the region $x \in (0,1)$.

\begin{figure}[!htbp]
\begin{center}
\includegraphics[height=.25\textheight]{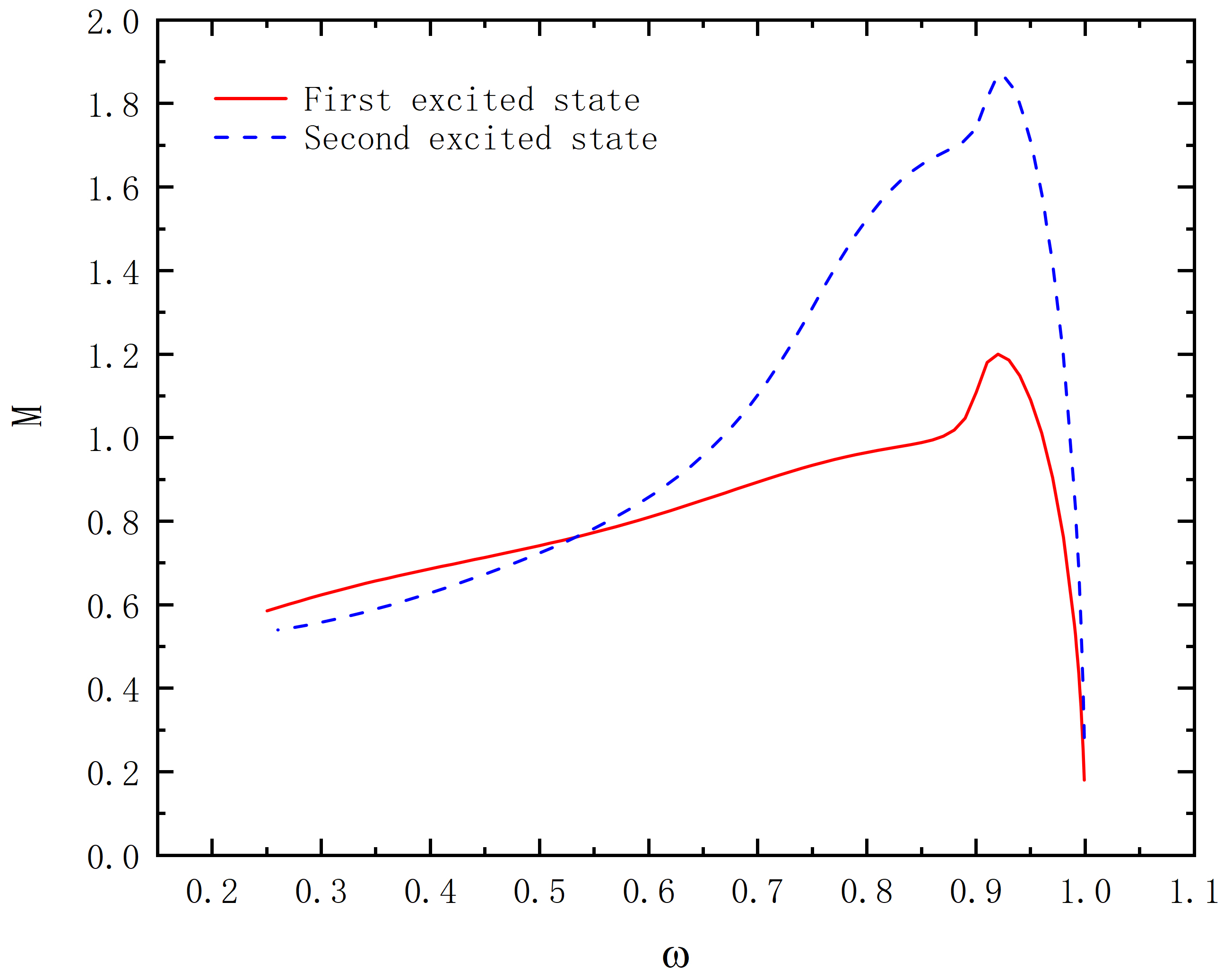}
\includegraphics[height=.25\textheight]{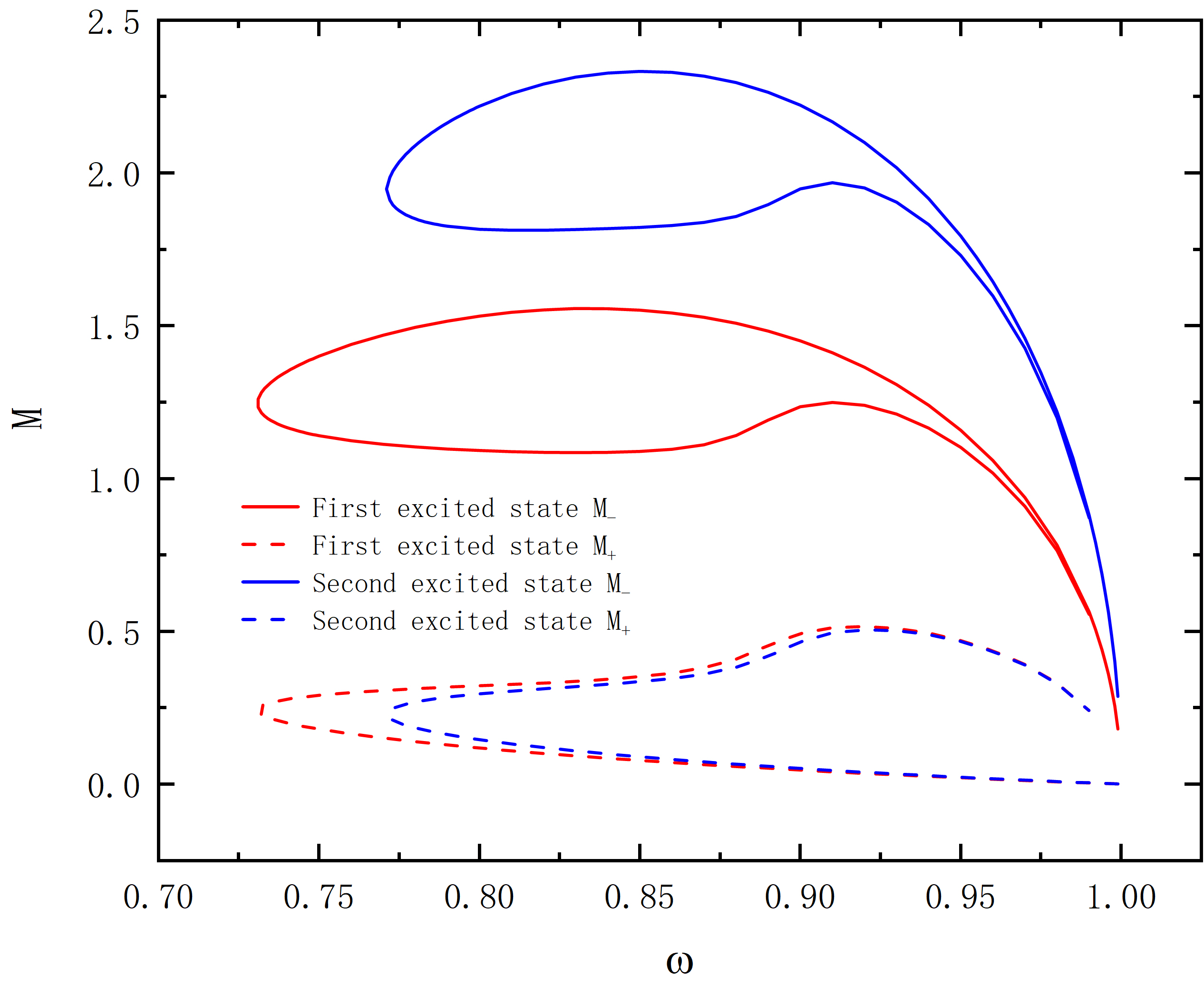}
\includegraphics[height=.25\textheight]{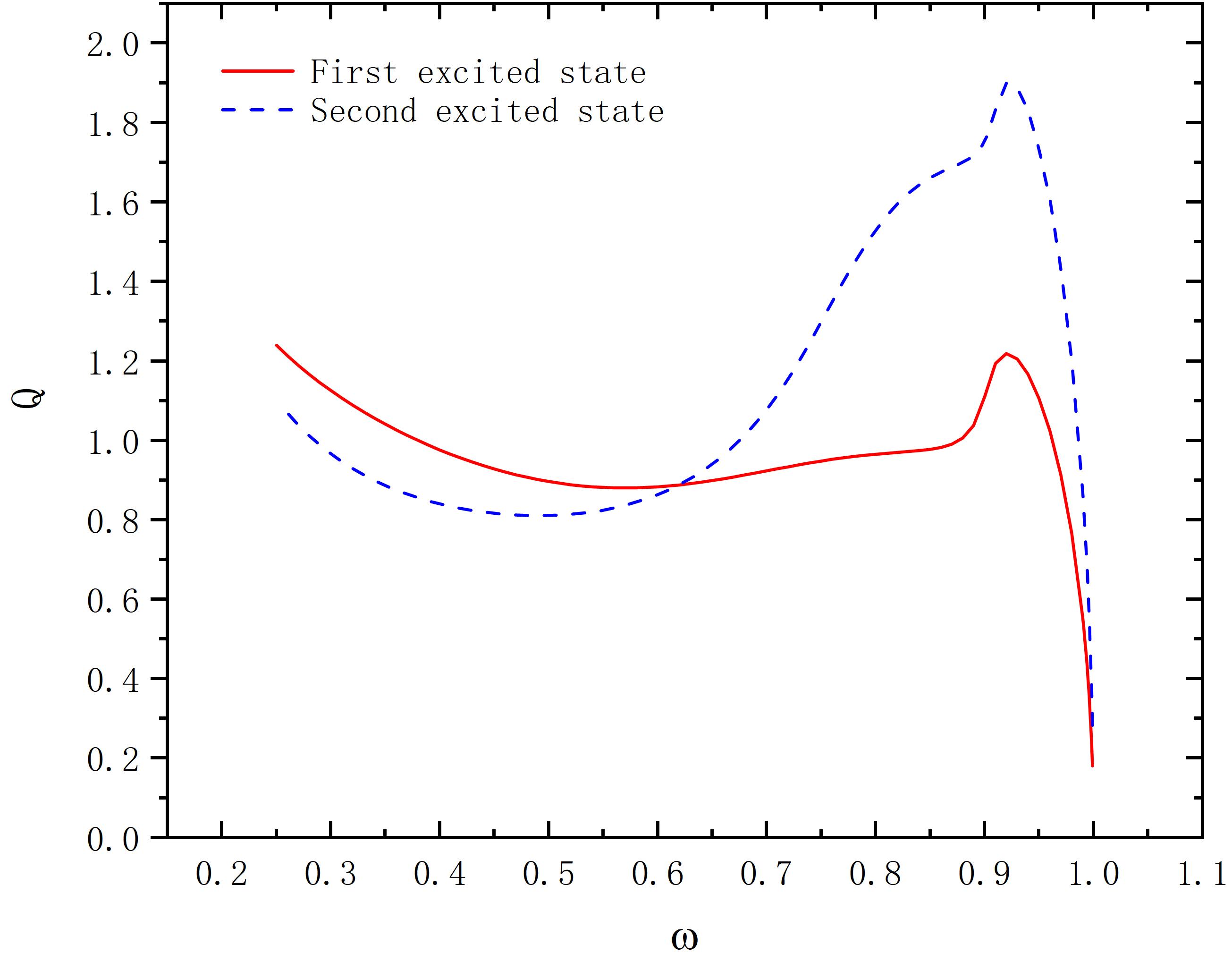}
\includegraphics[height=.25\textheight]{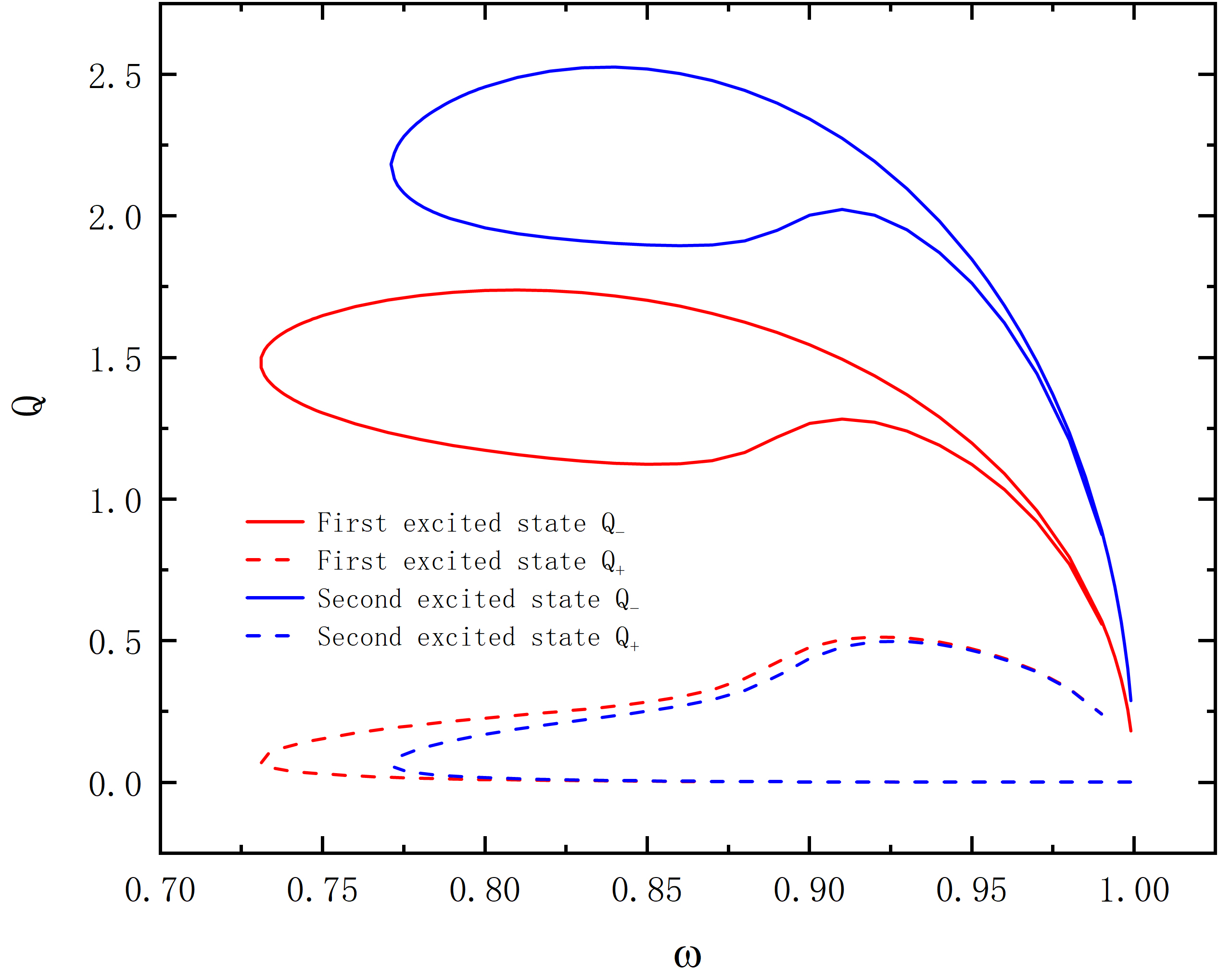}
\end{center}
\caption{The mass $M$ and the charge $Q$ as functions of frequency $\omega$ for $\eta_0=0.5$ in the case of symmetry and asymmetry.}
\label{asandsmq}
\end{figure}

In Fig.\ref{asandsmq}, the mass and noether charge for $\eta_0=0.5$ are exhibited. The figure indicates the excited state of superior level have higher magnitude of $M_-$ and $Q_-$. And liking ground state, the shape of curve $M$ resembles $Q$. Moreover, the symmetric solutions are $one-branch$, it's same to solution of ground state. Nevertheless, the asymmetrical solutions don't transform to the symmetric solutions, but rather presented as multi-valued function, furthermore, we discover the curve appear $multi-branch$ in all $\eta_0$ for excited state. This alteration is highly likely caused by zero points.

\subsection{The geometries}

\begin{figure}[!htbp]
\begin{center}
\includegraphics[height=.30\textheight]{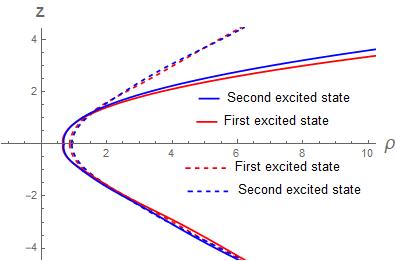}
\end{center}
\caption{Geometry of throats: isometric embedding in two dimensional view. Dashed lines represent symmetric solutions while solid lines correspond to asymmetrical solutions, all solutions have $\eta_0=0.5,\omega=0.8$.}
\label{EMB2}
\end{figure}

\begin{figure}[!htbp]
\begin{center}
\includegraphics[height=.22\textheight]{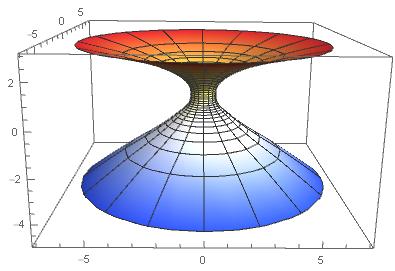}
\includegraphics[height=.22\textheight]{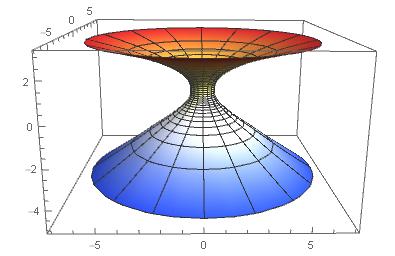}
\includegraphics[height=.26\textheight]{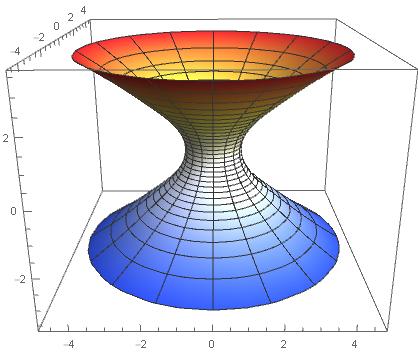}
\includegraphics[height=.26\textheight]{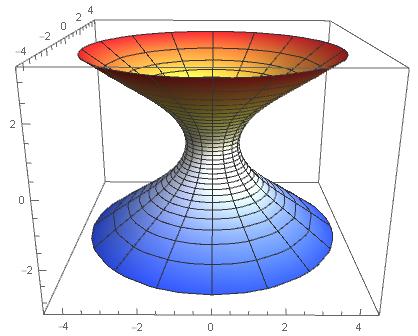}
\end{center}
\caption{Geometry of throat: isometric embedding in three dimensional view. Asymmetric solution for first excited state(upper left panel), second excited state(upper right panel), and symmetric solution for first excited state(below left panel), second excited state(below right panel), all solutions have $\eta_0=0.5,\omega=0.80$.}
\label{Geo3}
\end{figure}

The equatorial planes embedding in Euclidean space is presented in Fig.\ref{EMB2}, showcasing the characteristics of the first excited state and second excited state in both symmetric and asymmetric configurations for $\eta_0=0.5, \omega=0.8$. Moreover, we map three dimensional view in Fig.\ref{Geo3} to observe structure of throat. Surprisingly, transition of adjacent excited state have little effect on the geometry of the throat, indicating its stability in this alteration.

\section{Summary}\label{sec6}
In this paper, we construct the model of both symmetry and asymmetry in the system of wormhole embedded in a complex bosonic matter field and analyze the properties of this mixed system.

For ground state, we firstly separate solutions into the $c\neq 0$ and $c=0$ to study how different parameters influence the results. For the case of both symmetry and asymmetry, we find the effect of parameters $c$ and $\eta_0$ are similar in the curve shape, low value of $\eta_0$ and $c$ have the trend of generating $multi-branch$ solutions at restricted range of $\omega$ and when $\eta_0$ is big and $c \rightarrow 0$, the symmetric results develop into the scalar field solutions without phantom field in the two independent universes while the solutions of asymmetry degenerate to a single scalar field solution only in one asymptotically flat region.

Furthermore, in the discussion of transition between the asymmetry and symmetry, we present the variation of metric functions $A$, $p$ and field function $\phi$ as $\omega$ decreases at same $\eta_0=1$. These three functions all turn into the symmetric model under a threshold of $\omega$, which implies there is no asymmetric solution.

We also embed the equatorial plane of wormhole in Euclidean space and we discover the symmetric solutions of single throat will turn to double throats while the asymmetry barely becomes more pronounced as the $\omega$ decreasing. Moreover, we map the wormhole geometries in three dimensions to illustrate properties of throats.

Aiming to get more detailed conclusion, we study the solutions of excited state. Using the number of zero points, we divide the solutions into various level and first excited state and second excited state are concentrated. There are some resemblance between ground state and excited state, for instance, the curve of $M$ and $Q$ are similar and the symmetric solutions are $one-branch$ solutions when $\eta_0$ has a high value. In contrast, the asymmetric solutions don't demonstrate comparable behaviors, the transition between the asymmetry and symmetry not happen, instead, the results appear $multi-branch$ solutions for all value of $\eta_0$. We also plot the wormhole geometries, but there are not discernible disparity between Adjacent excited state.

In the future, we plan to expand the depth of our research. To begin with, the stability of these solutions will be explored. Next, considering there are only relationship between asymmetric solutions and symmetric solutions in this paper and solutions of parity-odd symmetry has been explored\cite{Yue:2023ela}, we will seek to study the link between parity-odd symmetry and asymmetry. What's more, we will try adding the number of field, such as appending a bosonic field in the first excited state in the basis of the mixed system in this paper and investigate the related property.

\section*{Acknowledgements}
YQW is supported by National Key Research and Development Program of China (Grant No. 2020YFC2201503)
and the National Natural Science Foundation of China (Grant No. 12275110 and 12247101)

\providecommand{\href}[2]{#2}\begingroup\raggedright
\endgroup


\begin{thebibliography}{10}

\bibitem{Wheeler:1955zz}
J.~A.~Wheeler,
``Geons,''
Phys. Rev. \textbf{97} (1955), 511-536
\bibitem{Power:1957zz}
E.~A.~Power and J.~A.~Wheeler,
``Thermal Geons,''
Rev. Mod. Phys. \textbf{29} (1957), 480-495
\bibitem{Kaup:1968zz}
D.~J.~Kaup,
``Klein-Gordon Geon,''
Phys. Rev. \textbf{172} (1968), 1331-1342
\bibitem{Ruffini:1969qy}
R.~Ruffini and S.~Bonazzola,
``Systems of selfgravitating particles in general relativity and the concept of an equation of state,''
Phys. Rev. \textbf{187} (1969), 1767-1783
\bibitem{Mielke:1996a}
Schunck, F.E., Mielke, E.W. (1996). Rotating Boson Stars. In: Hehl, F.W., Puntigam, R.A., Ruder, H. (eds) Relativity and Scientific Computing. Springer, Berlin, Heidelberg. $https://doi.org/10.1007/978-3-642-95732-1_ 7$
\bibitem{Abe:2010ap}
F.~Abe,
``Gravitational Microlensing by the Ellis Wormhole,''
Astrophys. J. \textbf{725} (2010), 787-793
doi:10.1088/0004-637X/725/1/787
[arXiv:1009.6084 [astro-ph.CO]].
\bibitem{Toki:2011zu}
Y.~Toki, T.~Kitamura, H.~Asada and F.~Abe,
``Astrometric Image Centroid Displacements due to Gravitational Microlensing by the Ellis Wormhole,''
Astrophys. J. \textbf{740} (2011), 121
doi:10.1088/0004-637X/740/2/121
[arXiv:1107.5374 [astro-ph.CO]].
\bibitem{Takahashi:2013jqa}
R.~Takahashi and H.~Asada,
``Observational Upper Bound on the Cosmic Abundances of Negative-mass Compact Objects and Ellis Wormholes from the Sloan Digital Sky Survey Quasar Lens Search,''
Astrophys. J. Lett. \textbf{768} (2013), L16
doi:10.1088/2041-8205/768/1/L16
[arXiv:1303.1301 [astro-ph.CO]].
\bibitem{Kanti:2011jz}
P.~Kanti, B.~Kleihaus and J.~Kunz,
``Wormholes in Dilatonic Einstein-Gauss-Bonnet Theory,''
Phys. Rev. Lett. \textbf{107} (2011), 271101
doi:10.1103/PhysRevLett.107.271101
[arXiv:1108.3003 [gr-qc]].
\bibitem{Kanti:2011yv}
P.~Kanti, B.~Kleihaus and J.~Kunz,
``Stable Lorentzian Wormholes in Dilatonic Einstein-Gauss-Bonnet Theory,''
Phys. Rev. D \textbf{85} (2012), 044007
doi:10.1103/PhysRevD.85.044007
[arXiv:1111.4049 [hep-th]].
\bibitem{Charalampidis:2013ixa}
E.~Charalampidis, T.~Ioannidou, B.~Kleihaus and J.~Kunz,
``Wormholes Threaded by Chiral Fields,''
Phys. Rev. D \textbf{87} (2013) no.8, 084069
doi:10.1103/PhysRevD.87.084069
[arXiv:1302.5560 [gr-qc]].
\bibitem{Dzhunushaliev:2022elv}
V.~Dzhunushaliev, V.~Folomeev, B.~Kleihaus and J.~Kunz,
``Mixed neutron-star-plus-wormhole systems: Rotating configurations,''
Phys. Rev. D \textbf{107} (2023) no.4, 044060
doi:10.1103/PhysRevD.107.044060
[arXiv:2210.04425 [gr-qc]].
\bibitem{Ellis:1973yv}
H.~G.~Ellis,
``Ether flow through a drainhole - a particle model in general relativity,''
J. Math. Phys. \textbf{14} (1973), 104-118
doi:10.1063/1.1666161
\bibitem{Ellis:1979bh}
H.~G.~Ellis,
``THE EVOLVING, FLOWLESS DRAIN HOLE: A NONGRAVITATING PARTICLE MODEL IN GENERAL RELATIVITY THEORY,''
Gen. Rel. Grav. \textbf{10} (1979), 105-123
doi:10.1007/BF00756794
\bibitem{Bronnikov:1973fh}
K.~A.~Bronnikov,
``Scalar-tensor theory and scalar charge,''
Acta Phys. Polon. B \textbf{4} (1973), 251-266
\bibitem{Morris:1988cz}
M.~S.~Morris and K.~S.~Thorne,
``Wormholes in space-time and their use for interstellar travel: A tool for teaching general relativity,''
Am. J. Phys. \textbf{56} (1988), 395-412
doi:10.1119/1.15620
\bibitem{Morris:1988tu}
M.~S.~Morris, K.~S.~Thorne and U.~Yurtsever,
``Wormholes, Time Machines, and the Weak Energy Condition,''
Phys. Rev. Lett. \textbf{61} (1988), 1446-1449
doi:10.1103/PhysRevLett.61.1446
\bibitem{Lobo:2005us}
F.~S.~N.~Lobo,
``Phantom energy traversable wormholes,''
Phys. Rev. D \textbf{71} (2005), 084011
doi:10.1103/PhysRevD.71.084011
[arXiv:gr-qc/0502099 [gr-qc]].
\bibitem{Sahni:1999qe}
V.~Sahni and L.~M.~Wang,
``A New cosmological model of quintessence and dark matter,''
Phys. Rev. D \textbf{62} (2000), 103517
[arXiv:astro-ph/9910097 [astro-ph]].
\bibitem{Matos:2000ng}
T.~Matos and L.~A.~Urena-Lopez,
``Quintessence and scalar dark matter in the universe,''
Class. Quant. Grav. \textbf{17} (2000), L75-L81
[arXiv:astro-ph/0004332 [astro-ph]].
\bibitem{Hu:2000ke}
W.~Hu, R.~Barkana and A.~Gruzinov,
``Cold and fuzzy dark matter,''
Phys. Rev. Lett. \textbf{85} (2000), 1158-1161
[arXiv:astro-ph/0003365 [astro-ph]].
\bibitem{Suarez:2013iw}
A.~Su\'arez, V.~H.~Robles and T.~Matos,
``A Review on the Scalar Field/Bose-Einstein Condensate Dark Matter Model,''
Astrophys. Space Sci. Proc. \textbf{38} (2014), 107-142
[arXiv:1302.0903 [astro-ph.CO]].
\bibitem{Hui:2016ltb}
L.~Hui, J.~P.~Ostriker, S.~Tremaine and E.~Witten,
``Ultralight scalars as cosmological dark matter,''
Phys. Rev. D \textbf{95} (2017) no.4, 043541
[arXiv:1610.08297 [astro-ph.CO]].
\bibitem{Padilla:2019fju}
L.~E.~Padilla, J.~A.~V\'azquez, T.~Matos and G.~Germ\'an,
``Scalar Field Dark Matter Spectator During Inflation: The Effect of Self-interaction,''
JCAP \textbf{05} (2019), 056
[arXiv:1901.00947 [astro-ph.CO]].
\bibitem{Colpi:1986ye}
M.~Colpi, S.~L.~Shapiro and I.~Wasserman,
``Boson Stars: Gravitational Equilibria of Selfinteracting Scalar Fields,''
Phys. Rev. Lett. \textbf{57} (1986), 2485-2488
\bibitem{Mielke:1980sa}
E.~W.~Mielke and R.~Scherzer,
``Geon Type Solutions of the Nonlinear {Heisenberg-Klein-Gordon} Equation,''
Phys. Rev. D \textbf{24} (1981), 2111
\bibitem{Kling:2017hjm}
F.~Kling and A.~Rajaraman,
``Profiles of boson stars with self-interactions,''
Phys. Rev. D \textbf{97} (2018) no.6, 063012
[arXiv:1712.06539 [hep-ph]].
\bibitem{Herdeiro:2020jzx}
C.~A.~R.~Herdeiro and E.~Radu,
``Asymptotically flat, spherical, self-interacting scalar, Dirac and Proca stars,''
Symmetry \textbf{12} (2020) no.12, 2032
[arXiv:2012.03595 [gr-qc]].
\bibitem{harrison2002numerical}
R.~Harrison, I.~Moroz and K.~P.~Tod,
``A numerical study of the Schrödinger–Newton equations,''
Nonlinearity \textbf{16} (2002) no.1, 101-122
\bibitem{Dzhunushaliev:2014bya}
V.~Dzhunushaliev, V.~Folomeev, C.~Hoffmann, B.~Kleihaus and J.~Kunz,
``Boson Stars with Nontrivial Topology,''
Phys. Rev. D \textbf{90} (2014) no.12, 124038
doi:10.1103/PhysRevD.90.124038
[arXiv:1409.6978 [gr-qc]].
\bibitem{Hoffmann:2017jfs}
C.~Hoffmann, T.~Ioannidou, S.~Kahlen, B.~Kleihaus and J.~Kunz,
``Spontaneous symmetry breaking in wormholes spacetimes with matter,''
Phys. Rev. D \textbf{95} (2017) no.8, 084010
doi:10.1103/PhysRevD.95.084010
[arXiv:1703.03344 [gr-qc]].
\bibitem{Hoffmann:2018oml}
C.~Hoffmann, T.~Ioannidou, S.~Kahlen, B.~Kleihaus and J.~Kunz,
``Symmetric and Asymmetric Wormholes Immersed In Rotating Matter,''
Phys. Rev. D \textbf{97} (2018) no.12, 124019
doi:10.1103/PhysRevD.97.124019
[arXiv:1803.11044 [gr-qc]].
\bibitem{Coleman:1985ki}
S.~R.~Coleman,
``Q-balls,''
Nucl. Phys. B \textbf{262} (1985) no.2, 263
doi:10.1016/0550-3213(86)90520-1
\bibitem{Wald:1984rg}
R.~M.~Wald,
``General Relativity,''
Chicago Univ. Pr., 1984,
doi:10.7208/chicago/9780226870373.001.0001
\bibitem{Yue:2023ela}
Y.~Yue, P.~B.~Ding and Y.~Q.~Wang,
``Boson star with parity-odd symmetry in wormhole spacetime,''
[arXiv:2305.04496 [gr-qc]].
\end{thebibliography}
\end{document}